\documentclass[paper]{JHEP}
\usepackage[centertags]{amsmath}
\usepackage{amsfonts}
\usepackage{amssymb}
\usepackage{amsthm}
\usepackage{graphicx}
\usepackage{psfrag}
\newcommand{\ket}[1]{|{#1}\rangle}
\def\one{{\rm 1\kern -.9mm l}}

\def\beq{\begin{equation}}
\def\eeq{\end{equation}}
\def\beqa{\begin{eqnarray}}
\def\eeqa{\end{eqnarray}}
\newcommand{\eq}[1]{eq. (\ref{#1})}
\newcommand{\comm}[2]{\left[#1\,,\,#2\right]}
\newcommand{\acomm}[2]{\left\{#1\,,\,#2\right\}}
\newcommand{\normord}[1]{:\! #1 \! : \!}
\newcommand{\Tr}{\mathrm{Tr}\,}

\def\ii{\mathrm{i}}
\def\ee{\mathrm{e}}

\newcommand{\lvev}{\langle\hskip -6pt\langle\hskip 4pt}
\newcommand{\rvev}{\hskip 4pt\rangle\hskip -6pt\rangle}
\newcommand{\bea}{\begin{eqnarray}}
\newcommand{\ena}{\end{eqnarray}}

\def\ii{\mathrm{i}}
\def\ee{\mathrm{e}}
\title{$\mathcal{N}=1/2$ quiver gauge theories from open strings with R-R fluxes
\thanks{Work partially supported by the European Community's Human
Potential Programme under contract MRTN-CT-2004-005104
``{Constituents, Fundamental Forces and Symmetries
of the Universe}'', and by the Italian M.I.U.R under contract
PRIN-2003023852 ``{Physics of fundamental interactions: gauge theories,
gravity and strings}''.}}

\author{
Marco Bill\'o
,
Marialuisa Frau
,
Fabio Lonegro
\\
Dipartimento di Fisica Teorica, Universit\`a di Torino\\
and Istituto Nazionale di Fisica Nucleare - sezione di Torino \\
via P. Giuria 1, I-10125 Torino, Italy
}
\author{
Alberto Lerda
\thanks{E-mails: \tt{billo,frau,lonegro,lerda@to.infn.it}}\\
Dipartimento di Scienze e Tecnologie Avanzate\\
Universit\`a del Piemonte Orientale\\
via V. Bellini 25/G, I-15100 Alessandria, Italy\\
and Istituto Nazionale di Fisica Nucleare  - sezione di Torino
}

\abstract{We consider a four dimensional $\mathcal{N}=1$ gauge theory
with bifundamental matter and a superpotential, defined on stacks of fractional
branes. By turning on a flux for the R-R graviphoton field strength and computing open string
amplitudes with insertions of R-R closed string vertices, we
introduce a non-anticommutative deformation and obtain the $\mathcal{N}=1/2$
version of the theory. We also comment on the appearance of a new structure in the effective
Lagrangian.}

\preprint{DFTT-01/2005}

\begin{document}

\section{Introduction}
\label{sec:intro}
The study of deformed gauge theories has recently attracted a lot
of interest, especially after it has become evident the connection
between field theory deformation parameters and non-trivial
geometric backgrounds. This connection is particularly clear in a string
theory context where the gauge theories describe the low-energy
dynamics of open strings attached to D-branes and the
deformation parameters are associated to non-trivial fluxes for some
closed string fields to which the D-branes can couple.
The most notable example of this relation is provided by the
non-commutative gauge theories which can be efficiently
described in terms of open strings propagating in a background
with a constant NS-NS $B_{\mu\nu}$ field \cite{Seiberg:1999vs}.
More recently, other types of backgrounds have been
considered by turning on fluxes for suitable combinations of the
anti-symmetric tensor fields of the closed string spectrum. Among
the various possibilities that have been explored, there is the
one in which a graviphoton background of the R-R sector is turned
on.

As explained in Refs.~
\cite{deBoer:2003dn,Ooguri:2003qp,Ooguri:2003tt,Seiberg:2003yz,Berkovits:2003kj},
a constant self-dual graviphoton
field strength $C_{\mu\nu}$ induces a deformation of the four dimensional
superspace \cite{Ferrara:2000mm,Klemm:2001yu}
in which the fermionic coordinates cease to be
anticommuting Grassmann variables and become elements of a Clifford
algebra, namely
\begin{equation}
\big\{\theta^{\alpha},\theta^{\beta}\big\} \,=\,C^{\alpha\beta}~~,~~
\big\{\theta^{\alpha},\bar\theta^{\dot\beta}\big\} \, = \,
\big\{\bar\theta^{\dot\alpha},\bar\theta^{\dot\beta}\big\} \,=\,0
\label{thetatheta}
\end{equation}
where
$C^{\alpha\beta}=\frac{1}{4}C_{\mu\nu}(\sigma^{\mu\nu})^{\alpha\beta}$.
The non-vanishing anticommutator in (\ref{thetatheta}) breaks the
four dimensional Lorentz group $\mathrm{SU}(2)_\mathrm{L}\times
\mathrm{SU}(2)_\mathrm{R}$ to $\mathrm{SU}(2)_\mathrm{L}$ and
reduces the number of conserved supercharges by a factor of two.
Therefore, a graviphoton background deforms a $\mathcal{N}=1$
field theory in four dimensions into a $\mathcal{N}=1/2$ theory
with only two supercharges. Furthermore, new types of interactions
and couplings are induced by the non-anticommutative structure of
superspace.

Supersymmetric field theories based on non-anticommutative superspaces
(which we will call simply non-anticommutative, or NAC, field theories) have
been the subject of vast investigation in the recent past from
many different points of view
%\cite{Britto:2003aj,Terashima:2003ri,Araki:2003se,Grisaru:2003fd,
%Britto:2003kg,Lunin:2003bm,Berenstein:2003sr,Alishahiha:2003kg,
%Imaanpur:2003jj,Grassi:2003qk,Britto:2003uv,Banin:2004hy,Jack:2004pq,
%Penati:2004gh,Azorkina:2005mx,Hatanaka:2005rg}.
[9--24].
In this paper, extending our previous work~\cite{Billo:2004zq},
we will analyze $\mathcal{N}=1/2$ gauge theories
with matter in the fundamental or bifundamental representation
working explicitly in a stringy set-up. In particular, we will
engineer a $\mathcal{N}=1$ gauge theory in four dimensions 
by considering stacks of
fractional D3-branes in the orbifold
$\mathbb{C}^3/(\mathbb{Z}_2\times\mathbb{Z}_2)$; the open strings
starting and ending on the same type of fractional D-branes
describe the gauge multiplets, while the strings stretching
between two different types of D-branes describe chiral and
anti-chiral matter multiplets in bifundamental representations.
We then demonstrate that a NAC deformation of this quiver gauge
theory, including its superpotential,
appears by turning on a graviphoton background with constant field
strength in the R-R sector. The presence of a non trivial R-R flux
modifies the dynamics of the open strings and introduces new
couplings that correspond to mixed open/closed string amplitudes
which we explicitly compute.
These new interactions are the same as those which can be derived from
the NAC deformation of superspace. However, we also find an extra
coupling which cannot be immediately obtained from the NAC
superspace.
Our approach provides in principle a unified way of treating
various deformations on gauge theories by computing mixed
open/closed string amplitudes, and shows that, at least when the
flux is constant, the NSR formulation of string theory
allows to treat also a R-R background.

This paper is organized as follows: in section \ref{sec:gauge}, using a
superspace approach, we first discuss the NAC deformation of a 
$\mathrm{U}(N)$ gauge theory with fundamental matter, and then 
the NAC structure of a quiver gauge theory with group 
$\mathrm{U}(N_0)\times\mathrm{U}(N_1)\times\mathrm{U}(N_2)\times\mathrm{U}(N_3)$
and matter in bifundamental representations.
In section \ref{sec:frac} we show how to engineer the above quiver
theory, including its superpotential,
with fractional D3 branes of Type IIB string theory in the
orbifold $\mathbb{C}^3/(\mathbb{Z}_2\times\mathbb{Z}_2)$, while in
section \ref{sec:def} we explicitly derive the deformation induced
by a R-R graviphoton background on the massless open string dynamics by computing
mixed open/closed string amplitudes in the NSR formalism.
Finally, in appendix \ref{appendix} we list our conventions and collect some
technical details that are useful to reproduce our calculations.

\section{${\cal N}=1/2$ gauge theories with fundamental matter}
\label{sec:gauge}
In this section we review the NAC deformation of ${\cal N}=1$
gauge theories; we use a superspace approach first to describe deformed
superfields, and then to introduce gauge invariant actions for NAC theories
with chiral matter in fundamental or bifundamental representations.

\subsection{Superfields in NAC superspace}
In terms of the commuting chiral coordinates $y^\mu \equiv x^\mu + \ii\,
\theta{\sigma}^\mu \bar\theta$, a vector superfield $V$ in the WZ gauge has the
following expansion~\footnote{Our Euclidean
conventions are given in appendix \ref{subsec:conventions}.}
\begin{equation}
V~=~
-2\,\theta\sigma^\mu\bar\theta\,A_\mu(y)
-2\ii \,\bar\theta\bar\theta\,\theta\lambda(y)
-2\ii \,\theta\theta\,\bar\theta\,\overline\lambda(y)
+\theta\theta\,\bar\theta\bar\theta\,\big(\ii\,\partial\cdot
A(y)+D(y)\big)
\label{V}
\end{equation}
where $A_\mu$ is the gauge vector field, $\lambda$ and
$\overline\lambda$ are the gauginos and $D$ is an auxiliary
field. Clearly these components transform in the adjoint
representation of the gauge group, and
the residual transformations which preserve the WZ
gauge are of the form
\begin{equation}
\ee^V ~\rightarrow ~\ee^{V'} = \ee^{-\ii\,\overline\Xi}\,\ee^V\,\ee^{\ii\,\Xi}
\label{gaugetransf}
\end{equation}
with $\Xi$ and $\overline\Xi$ given by
\begin{equation}
\Xi ~= ~\varepsilon(y)~~~~,~~~~
\overline \Xi ~ = ~\varepsilon(y)
-2\ii\,\theta\sigma^\mu\bar\theta\,
\partial_\mu\varepsilon(y)
-\theta\theta\,\bar\theta\bar\theta\,\partial^2\varepsilon(y)
\label{Xi}
\end{equation}
in terms of the gauge parameter $\varepsilon(y)$. Indeed, by expanding
(\ref{gaugetransf})
in $\theta$ and $\bar \theta$, one can easily find the standard infinitesimal
gauge transformations for the components
\begin{equation}
\begin{aligned}
\delta A_\mu ~& = ~ \partial_\mu \varepsilon
+\ii\,\comm{A_\mu}{\varepsilon}~~~,~~~
\delta \lambda ~ = ~\ii\,\comm{\lambda}{\varepsilon}\\
%~~~,~~~
\delta \overline\lambda ~& = ~\ii\,\comm{\overline\lambda}{\varepsilon}
~~~,~~~
\delta D~ = ~\ii\,\comm{D}{\varepsilon}~~.
\end{aligned}
\label{gaugetransf1}
\end{equation}

On the other hand, a chiral superfield $\Phi$ has the following $\theta$-expansion
\begin{equation}
\Phi = \varphi(y)+\sqrt{2}\,\theta
\chi(y)+\theta\theta\,F(y)
\label{Phi}
\end{equation}
where $\varphi$ is a complex scalar field, $\chi$ is a chiralino
and $F$ is an auxiliary field.
Correspondingly, an anti-chiral superfield $\overline\Phi$ is given by
\begin{equation}
\begin{aligned}
\overline\Phi ~=~ &\overline\varphi(y) - 2\ii
\,\theta\sigma^\mu\bar\theta\,\partial_\mu\overline\varphi(y)-
\theta\theta\,\bar\theta\bar\theta\,
\partial^2\overline\varphi(y)
\\
& +\sqrt{2}\,\bar\theta\overline\chi(y)
-2\sqrt{2}\ii\,\theta\sigma^\mu\bar\theta\,\bar\theta\partial_\mu\overline\chi(y)
+\bar\theta\bar\theta\, \overline F(y)
\end{aligned}
\label{barPhi}
\end{equation}
in terms of the conjugate components. When these fields are in the fundamental
and anti-fundamental representations
of the gauge group, their gauge transformations
are
\begin{equation}
\Phi ~\rightarrow ~\Phi'\,=\,\ee^{-\ii\,\Xi}\,\Phi~~~~,~~~~
\overline\Phi  ~\rightarrow ~{\overline
\Phi}'\,=\,\overline\Phi~\ee^{\ii\,\overline\Xi}~~,
\label{gaugetransf2}
\end{equation}
which imply the following infinitesimal transformations for the
components
\begin{equation}
\label{gaugetransf3}
\begin{aligned}
\delta\varphi \,&=\,-\ii\,\varepsilon\,\varphi~~~,~~~
\delta \chi \,= \,-\ii\,\varepsilon\,\chi~~~,~~~
\delta F\,= \,-\ii\,\varepsilon\,F~~,\\
\delta\overline\varphi \,&=\,\ii\,\overline\varphi\,\varepsilon~~~~~,~~~
\delta \overline\chi \,= \,\ii\,\overline\chi\,\varepsilon~~~~~,~~~
\delta \overline F\,= \,\ii\,\overline F\,\varepsilon~~.
\end{aligned}
\end{equation}

In Ref.~\cite{Seiberg:2003yz} the consequences of the NAC
deformation (\ref{thetatheta}) of the superspace have been
analyzed and interpreted. First of all, in the presence of $C^{\alpha\beta}$
a new product among superfields, called $\star\,$-product, must be introduced according to
\begin{equation}
\begin{aligned}
\Psi_1\,\star\,\Psi_2 \,&=\,
\Psi_1\,\exp\!\left(
-\frac{C^{\alpha\beta}}{2}\,\overleftarrow{\frac{\partial}{\partial\theta^\alpha}}
\,\overrightarrow{\frac{\partial}{\partial\theta^\beta}}
\right)\Psi_2
\\
&=\,
\Psi_1\,\Psi_2 - C^{\alpha\beta}\!
\left(\psi_{1\alpha}+\sqrt{2}\,\theta_\alpha f_1\right)\!
\left(\psi_{2\beta}+\sqrt{2}\,\theta_\beta f_2\right)
-\det C\,f_1\,f_2
\end{aligned}
\label{starproduct}
\end{equation}
where $\Psi_1$ and $\Psi_2$ are two arbitrary superfields,
$\psi_\alpha$ and $f$ are, respectively, their $\theta^\alpha$
and the $\theta\theta$ components (which in general can be
functions both of $y$ and of $\bar \theta$), and $\det C =\frac{1}{2}\,
C^{\alpha\beta}C_{\alpha\beta}
=\frac{1}{4}\,C^{\mu\nu}C_{\mu\nu}$.
Then, the parameterization (\ref{V}) of the vector superfield $V$
must be modified by shifting the gaugino $\lambda_\alpha$
according to
\begin{equation}
\lambda_\alpha ~\rightarrow ~
\lambda_\alpha-\frac{1}{2}
\,C_\alpha^{~\beta}
\sigma^\mu_{\beta\dot\alpha}\acomm{\overline\lambda^{\dot\alpha}}{A_\mu}
\label{newV}
\end{equation}
in such a way that the {\it standard} gauge transformations
(\ref{gaugetransf1}) can be derived from the $\star\,$-product
version of (\ref{gaugetransf}), {\it i.e.}
\begin{equation}
\ee^V ~\rightarrow ~\ee^{V'} =
\ee^{-\ii\,\overline\Xi}\,\star\,\ee^V\,\star\,\ee^{\ii\,\Xi}~~.
\label{newgaugetransf}
\end{equation}
In these expressions the exponentials are defined with the $\star\,$-product
and the gauge parameters are given by
\begin{equation}
\begin{aligned}
\Xi ~& = ~\varepsilon(y) \\
\overline \Xi ~& = ~\varepsilon(y)
-2\ii\,\theta\sigma^\mu\bar\theta\,
\partial_\mu\varepsilon(y)
-\theta\theta\,\bar\theta\bar\theta\,\partial^2\varepsilon(y)
+\ii\,\bar\theta\bar\theta
\,C^{\mu\nu}\,\acomm{\partial_\mu\varepsilon(y)}{A_\nu}~~.
\end{aligned}
\label{newXi}
\end{equation}
Note the appearance in $\overline\Xi$ of a $C$-dependent term that involves also the gauge
field.
Furthermore, from the deformed vector superfield one can obtain a deformed
field strength superfield ${\cal W}_\alpha$ by replacing ordinary
products with $\star\,$-products in the usual definition~\cite{Seiberg:2003yz}, {\it
i.e.}
\begin{equation}
{\cal W}_\alpha = -\frac{1}{8}\,{\overline D}^2\,\star\,\ee^{-V}\,\star\,D_\alpha\,\star\,\ee^{V}
\label{Walpha}
\end{equation}
where $D_\alpha$ and ${\overline D}_{\dot\alpha}$ are the standard
covariant derivatives. In this way one finds that
${\cal W}_\alpha$ acquires a deformation term proportional to
$C_\alpha^{~\beta}\,\theta_\beta\,{\overline
\lambda\,\overline \lambda}$.

This reasoning can be extended also to chiral and anti-chiral
superfields (see for example Ref.~\cite{Araki:2003se}). If one requires that the {\it standard}
gauge transformations of the matter fields (\ref{gaugetransf3})
follow from the $\star\,$-product version of
(\ref{gaugetransf2}), {\it i.e.} from
\begin{equation}
\Phi  ~\rightarrow ~\Phi'\,=\,\ee^{-\ii\,\Xi}\,\star\,\Phi
~~~~,~~~~
\overline\Phi  ~\rightarrow ~{\overline
\Phi}'\,=\,\overline\Phi~\star\,\ee^{\ii\,\overline\Xi}~~,
\label{newgaugetransf2}
\end{equation}
then the usual expansion (\ref{Phi}) of the chiral superfield can be kept, but
the parameterization of $\overline\Phi$ must be changed by
replacing in (\ref{barPhi}) the auxiliary field according to
\begin{equation}
\overline F ~\rightarrow ~\overline F + 2\ii\,C^{\mu\nu}\,
\partial_\mu\big(\overline\varphi\,A_\nu\big) -C^{\mu\nu}\,
\overline\varphi\,A_\mu\,A_\nu +\ii\,{a}\,C^{\mu\nu}\,\overline\varphi\,F_{\mu\nu}
+b\,\det C\,\overline\varphi\,\overline\lambda\,\overline\lambda
\label{F'}
\end{equation}
where $a$ and $b$ are free parameters. In Ref.~\cite{Araki:2003se} the
minimal choice $a=b=0$ was made but other choices are equally acceptable.
In any case, it is interesting to note that the $C$-deformation of
superspace induces in the anti-chiral superfield $\overline\Phi$ the appearance
of terms that depend on the gauge vector $A_\mu$ and possibly also on
the gaugino $\overline \lambda$.

\subsection{$\mathrm{U}(N)$ gauge theories with matter in NAC superspace}
\label{subsec:gaugetheories}
After these preliminaries, it is quite easy to write gauge invariant
actions for NAC theories with chiral matter in the fundamental representation.

Let us first consider the simplest example of a theory with gauge group
$\mathrm{U}(N)$. In this case the pure Yang-Mills part
of the Lagrangian is given by
\begin{equation}
{\cal L}_{\rm gauge} = \frac{\ii}{8\pi}\Bigg[
\int\!\! d^2\theta ~\tau\,{\rm
Tr}\Big({\cal W}\,\star\,{\cal W}\Big)
-\int\!\! d^2\bar\theta ~\bar\tau\,{\rm Tr}\Big(\overline{\cal W}\,\star\,\overline
{\cal W}\Big)\Bigg]
\label{ymlagrangian}
\end{equation}
where $\tau= \frac{\theta_{\rm
YM}}{2\pi}+\ii\,\frac{4\pi}{g^2}$ is the
complexified Yang-Mills coupling, and ${\cal W}$ is the deformed field strength
superfield (\ref{Walpha}). Expanding
(\ref{ymlagrangian}) in components, we find
\begin{equation}
\begin{aligned}
{\cal L}_{\rm gauge} ~=~&
\frac{1}{g^2}\,{\rm
Tr}\Big\{\frac{1}{2}\,F_{\mu\nu}^2 -2\ii\,\bar\lambda\,\bar\sigma^\mu
D_\mu\lambda - D^2
+2\ii\,C^{\mu\nu}\, F_{\mu\nu}\,
\overline\lambda\,\overline\lambda \\&-
4\,\det C\, \big(\,\overline\lambda\,\overline\lambda\,\big)^2\Big\}
-\,\frac{\ii\,\theta_{\rm YM}}{32\pi^2}\,\varepsilon^{\mu\nu\rho\sigma}\,{\rm
Tr}\,F_{\mu\nu}\,F_{\rho\sigma}
\end{aligned}
\label{ymlagrangian'}
\end{equation}
Note that the NAC deformation does not affect the $\theta_{\rm
YM}$-term which remains purely topological.
The action (\ref{ymlagrangian'}), which was first
written in Ref.~\cite{Seiberg:2003yz},
can also be obtained by computing scattering amplitudes of open
strings in a R-R graviphoton background as shown in Ref.~\cite{Billo:2004zq}.

The matter part of the Lagrangian is given by the usual expression
in which ordinary products are replaced by $\star\,$-products,
namely
\begin{equation}
{\cal L}_{\rm matt} = \int\!\! d^2\theta\,
d^2\bar\theta\,\,
\,\Big(\overline \Phi\,\star\,\ee^{V}\,\star\,\Phi\Big)
\label{mattlagrangian}
\end{equation}
The gauge invariance of ${\cal L}_{\rm matt}$
is manifest from the transformation
properties (\ref{newgaugetransf}) and (\ref{newgaugetransf2}), and
its explicit component form can be obtained with a straightforward
calculation that leads, modulo total derivative terms, to
\begin{equation}
\begin{aligned}
{\cal L}_{\rm matt} ~=~&D^\mu\overline\varphi\,D_\mu\varphi
-\ii\,\overline\chi\,\bar\sigma^\mu D_\mu\chi
+\overline F\,F +\overline\varphi\,D\,\varphi
+\sqrt 2\,{\ii}\,\big(\overline\chi\,\overline\lambda\,\varphi+
\overline\varphi\,\lambda\,\chi\big)
\\&~+\sqrt{2}\,C^{\mu\nu}\,
D_\mu\overline\varphi\,\overline\lambda\,\bar\sigma_\nu
\,\chi
+ \ii a'\,C^{\mu\nu}\,\overline\varphi\,F_{\mu\nu}\,F
+b'\det C\,\,\overline\varphi\,\overline\lambda\,\overline\lambda\,F
\end{aligned}
\label{mattlagrangian'}
\end{equation}
where $a'=a+1$ and $b'=b-1$ in terms of the parameters appearing in (\ref{F'}).
This Lagrangian was first
introduced and analyzed in Ref.~\cite{Araki:2003se} where, however, different
conventions were used and the choice $a'=-b'=1$ was made.

It can be shown in full generality that the complete
system $\Big({\cal L}_{\rm gauge}+{\cal L}_{\rm matt}\Big)$ is invariant,
up to total derivatives, only under a half of the original ${\cal N}=1$ supersymmetry
transformations, namely under
\begin{equation}
\begin{aligned}
\delta A_\mu&= \ii \,\xi \,\sigma_\mu \,\overline\lambda~~~~~~,~~~~~~\delta D= \xi\,
\sigma^\mu D_\mu\overline \lambda~~~~~~,~~~~~~
\delta \overline\lambda=0~~~~~~,\\
\delta \lambda&= \ii\,\xi\,D -\,\frac{1}{2}\,\xi\,\sigma^{\mu\nu}
 \Big(F_{\mu\nu}+{\ii}\,
C_{\mu\nu}\,\overline \lambda \,\overline\lambda+\ii\,
\frac{g^2}{2}\,C_{\mu\nu}\,F\,\overline\varphi\Big)
~~~~~~,\\
\delta \varphi &= \sqrt{2}\,\xi\,\chi ~~~~~~,~~~~~~\delta \overline\varphi =0 ~~~~~~,~~~~~~
\delta \chi = \sqrt{2}\,\xi\, F ~~~~~~,\\
\delta \overline\chi&=
-\sqrt{2}\ii\,D_\mu\overline\varphi\,\xi\,\sigma^\mu ~~~~~~,~~~~~~
\delta F=0~~~~~~,
\\
\delta \overline F&=\sqrt{2}\ii\,\xi\,\sigma^\mu\,
D_\mu\overline\chi
-2\ii \,\overline \varphi\,
\xi\,\lambda
+C^{\mu\nu}\Big(2\,D_\mu\overline\varphi\,\xi\,\sigma_\nu\,\overline\lambda+
(2a'-1)\,\overline\varphi\,\xi\,\sigma_\nu\,D_\mu\overline\lambda\\
&~~~~~~~~~~~~~~~~~~~~~~~~-\,\frac{\sqrt{2}}{4}g^2\,
\overline\varphi\,\xi\,\sigma_{\mu\nu}\chi\,\overline\varphi
\Big)\\
\end{aligned}
\label{susy}
\end{equation}
where $\xi$ is the chiral anti-commuting parameter. Notice
the presence of $C$-dependent terms proportional to the coupling constant
$g^2$ in the transformation laws of the gaugino $\lambda$ and the
auxiliary field $\overline F$, which were not previously
considered. The remaining
supersymmetries, associated to the anti-chiral parameter $\overline
\xi$, are explicitly broken by the NAC deformation.

The theory described by (\ref{mattlagrangian'})
can be regarded as the gauged version of the ${\cal N}=1/2$ Wess-Zumino model
whose renormalization properties have been recently studied
in the literature (see for example Refs.~\cite{Grisaru:2003fd}).
Due to the charge carried by the chiral superfield,
there is no room in (\ref{mattlagrangian'}) for a superpotential
term, and so if we want to investigate superpotentials we have to consider
a suitable extension of this theory, which we will do in the next subsection.
However, it is interesting to observe that in the present context 
it is possible to
introduce a supersymmetric and gauge invariant interaction term of the form
\begin{equation}
{\cal L}_{\rm int} = c'\,g^2\,\det C\,(\overline\varphi\,F)^2
\label{interaction}
\end{equation}
where $c'$ is a free parameter~\footnote{Terms like, for example,
$(\det C)^4\,(\overline\varphi\,F)^4$
or $(\det C)^4\,(\overline\varphi\,\overline\lambda\overline\lambda\,F)^2$
will not be considered since they explicitly break the
$\mathrm U(1)$ R-symmetry of the
theory.}. Such a term is compatible with the
$\star$-product structure of the model since it can be generated
by adding in (\ref{F'}) a further shift for the auxiliary field $\overline
F$ that respects all requirements, {\it i.e.}
\begin{equation}
\overline F~\rightarrow ~\overline F + c'\,g^2\,\det
C\,\overline\varphi\,F\,\overline\varphi
\label{F''}
\end{equation}
The interaction (\ref{interaction}), which survives also in the ungauged
theory, is not usually included 
in the Lagrangian of the
${\cal N}=1/2$ Wess-Zumino model, since in this case, using the equation
of motion for the auxiliary field, it becomes proportional to 
$\det C\,F^3$, {\it i.e.} to a term of the deformed Wess-Zumino
superpotential. However, in Ref.~\cite{Grisaru:2003fd} it has
been shown that a term precisely like (\ref{interaction}) appears 
in the 1-loop divergences of the ${\cal N}=1/2$ Wess-Zumino model.
In section \ref{sec:def} we will show that
an interaction of the form (\ref{interaction}) naturally appears in the
string realization of the NAC theories provided by D3 branes in
a R-R graviphoton background.

\subsection{Quiver gauge theories in NAC superspace}
\label{subsec:quiver_NAC}
We now generalize the above NAC construction
to the ${\cal N}=1$ quiver theory  with gauge group
$\mathrm{U}(N_0)\times\mathrm{U}(N_1)\times
\mathrm{U}(N_2)\times\mathrm{U}(N_3)$ which
has a natural realization  as the world-volume theory on a superposition of
fractional D-branes in the orbifold $\mathbb{C}^3/(\mathbb{Z}_2\times
\mathbb{Z}_2)$.
The field content of this model is summarized in the
quiver diagram of Fig. \ref{fig:quiver} and consists of four
vector multiplets $V^I$ ($I=0,1,2,3$), one for each factor of the gauge group,
and twelve chiral multiplets $\Phi^{IJ}$ (with $I\not = J$)
that transform in the bifundamental representation $(N_I,\overline N_J)$
of the $\mathrm{U}(N_I)\times\mathrm{U}(N_J)$ sub-group, together with the corresponding
anti-chiral multiplets $\overline\Phi^{JI}$ that transform
in the $(\overline N_I, N_J)$ representation.
\FIGURE{\centerline{
\psfrag{R0}{\small $R_0$}
\psfrag{R1}{\small $R_1$}
\psfrag{R2}{\small $R_2$}
\psfrag{R3}{\small $R_3$}
\psfrag{R0w}{\small $R_0$}
\psfrag{R1w}{\small $R_1$}
\psfrag{R2w}{\small $R_2$}
\psfrag{R3w}{\small $R_3$}
\includegraphics{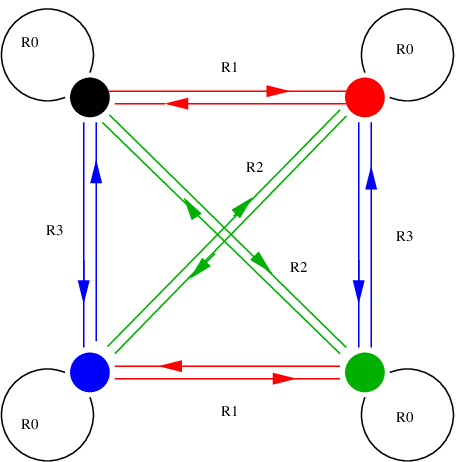}
\caption{\small{A quiver diagram encodes the field content and the charges of
a system of matter-coupled gauge theories. Each dot (labeled by $I=0,1,2,3$ in
this case) corresponds to a $\mathrm{U}(N_I)$ gauge group, for which a gauge
multiplet is considered. An oriented link from the $I$-th to the $J$-th dot
corresponds to a chiral multiplet $\Phi^{IJ}$ transforming in the
$(N_I,\overline N_J)$ representation.
As we will discuss in section \ref{sec:frac}, from a string theory point of view
this particular quiver diagram describes a system of fractional D-branes
of type II B in the orbifold
$\mathbb{C}^3/(\mathbb{Z}_2\times\mathbb{Z}_2)$. In this context
the dots represent the fractional branes and the lines the strings stretching
between them. For each type of string we indicate the representation of the
orbifold group in which the vertex operators should transform in order to
survive the orbifold projection.}}
\label{fig:quiver}
}}

A NAC deformation of the superspace induces several changes in this quiver
theory, which we now analyze. First of all, since the chiral and anti-chiral
superfields are in bifundamental representations, the transformation rules
(\ref{newgaugetransf2}) must be replaced by \begin{equation}
\begin{aligned}
\Phi^{IJ}  ~&\rightarrow ~{\Phi'}^{IJ}\,=\,\ee^{-\ii\,\Xi_I}\,\star\,
\Phi^{IJ}
\,\star\,\ee^{\ii\,\Xi_J}
\\
\overline\Phi^{JI}  ~&\rightarrow ~\overline
{\Phi'}^{JI}\,=\,\ee^{-\ii\,\overline\Xi_J}\,\star\,
\overline\Phi^{JI}~\star\,\ee^{\ii\,\overline\Xi_I}
\end{aligned}
\label{newgaugetransf3}
\end{equation}
where $\Xi_I$ and $\overline\Xi_J$ are
defined as in (\ref{Xi}). Then, if we require that
these formulas account for the appropriate gauge transformations of
the components, it is necessary to parameterize the superfields
as follows
\begin{equation}
\begin{aligned}
\Phi^{IJ} ~=~ &\varphi^{IJ}(y)+\sqrt{2}\,\theta
\chi^{IJ}(y)+\theta\theta\,F^{IJ}(y)
\\
\overline\Phi^{JI} ~=~ &\overline\varphi^{JI}(y) - 2\ii
\,\theta\sigma^\mu\bar\theta\,\partial_\mu\overline\varphi^{JI}(y)-
\theta\theta\,\bar\theta\bar\theta\,
\partial^2\overline\varphi^{JI}(y)
\\
~& +\sqrt{2}\,\bar\theta \overline\chi^{JI}(y)
-2\sqrt{2}\ii\,\theta\sigma^\mu\bar\theta\,\partial_\mu\overline\chi^{JI}(y)
+\bar\theta\bar\theta\, {\widetilde F}^{JI}(y)
\end{aligned}
\label{newPhi}
\end{equation}
where ${\widetilde F}^{JI} $ is given by the obvious
generalization of (\ref{F'}), {\it i.e.}
\begin{equation}
\begin{aligned}
{\widetilde F}^{JI} ~=~& \overline F^{JI} + 2\ii\,
C^{\mu\nu}\,\partial_\mu\big(\overline\varphi^{JI}\,A_{\nu}^I
+A_\nu^J\,\overline\varphi^{JI}\big)
-\ii\, C^{\mu\nu}\,\big(\overline\varphi^{JI}\,A_\mu^I A_{\nu}^I
+A_\mu^J A_{\nu}^J\,\overline\varphi^{JI}\big)
\\
~&+
\ii\,{a}\,C^{\mu\nu}\,\big(\overline\varphi^{JI}\,F_{\mu\nu}^I+
F_{\mu\nu}^J\,\overline\varphi^{JI}\big)+
b\,\det
C\,\big(\overline\varphi^{JI}\,\overline\lambda^I\,\overline\lambda^I
+\overline\lambda^J\,\overline\lambda^J\overline\varphi^{JI}\big)\\
~&
-2\,C^{\mu\nu}\, A_{\mu}^J\,
\overline\varphi^{JI}\, A_{\nu}^I
\end{aligned}
\label{tildeF}
\end{equation}
with $\overline F^{JI}$ being the auxiliary field conjugate to $F^{IJ}$.

The gauge invariant kinetic Lagrangian for a quiver theory in the
C-deformed superspace is simply given by
\begin{equation}
\label{Lquiver}
\begin{aligned}
{\cal L}_{K} =&\frac{\ii}{8\pi}\sum_I\Bigg[
\int\!\! d^2\theta ~\tau\,{\rm
Tr}\Big({\cal W}^I\star{\cal W}^I\Big)
-\int\!\! d^2\bar\theta ~\bar\tau\,{\rm Tr}\Big(\overline {\cal W}^I\star\overline
{\cal W}^I\Big)\Bigg]\\
&+\sum_{I\not = J}\int\!\! d^2\theta\,
d^2\bar\theta\,\,
\,{\rm Tr}\Big(\overline \Phi^{JI}\star\ee^{V^I}\star\Phi^{IJ}\star\ee^{-V^J}\Big)
\end{aligned}
\end{equation}
where ${\cal W}^I$ is the field-strength
superfield for the $I$-th node of the quiver diagram. Working out
the $\star$-products and expanding in $\theta$, after a lengthy but straightforward
calculation, one can find the component form of ${\cal L}_{K}$,
which turns out to be a natural generalization of what we
presented in the previous sub-section for the $\mathrm{U}(N)$ theory.
It is important to realize that in the quiver case we can add to the
Lagrangian also a gauge invariant superpotential term given by
\begin{equation}
{\cal L}_W+{\cal L}_{\overline W} = \frac{g}{3}\!
\sum_{I\not = J \not = K}\!\Bigg[\!\int\!\! d^2\theta\,
{\rm Tr}\Big(\Phi^{IJ}\star\Phi^{JK}\star\Phi^{KI}\Big)
+\int\!\! d^2\bar\theta\,
{\rm
Tr}\Big(\overline\Phi^{IJ}\star\overline\Phi^{JK}\star\overline\Phi^{KI}\Big)\Bigg]
\label{superpotentialterms}
\end{equation}
where the sum over the triples $I\not =J\not =K$ describes in fact
a sum over all possible triangles of the diagram in Fig. \ref{fig:quiver}
and the factor of $1/3$ eliminates the overcounting of cyclically symmetric terms.
The $\star$-products are easily evaluated in the holomorphic part
${\cal L}_W$, whose component form is
\begin{equation}
\begin{aligned}
{\cal L}_W \,=& \,\,g\!\sum_{I\not = J \not = K}\!{\rm
Tr}\Big(
F^{IJ}\varphi^{JK}\varphi^{KI} -
\varphi^{IJ}\chi^{JK}\chi^{KI}\Big)\\
&+\,g\!\sum_{I\not = J \not = K}\!{\rm
Tr}\Big(\frac{1}{4}\,
C^{\mu\nu}\,F^{IJ}\chi^{JK}\sigma^{\mu\nu}\chi^{KI}
-\frac{1}{3}\,\det C\,F^{IJ}F^{JK}F^{KI}
\Big)
\end{aligned}
\label{holsuppot}
\end{equation}
The anti-holomorphic piece ${\cal L}_{\overline W}$ is instead much more involved due to
the non-trivial parameterization of the anti-chiral superfields
$\overline\Phi^{IJ}$ given in (\ref{newPhi}) and
(\ref{tildeF}). Finding its complete component expression is
just a matter of lengthy algebra; however it is not difficult to
see that, among the many $C$-dependent terms, ${\cal L}_{\overline W}$
contains the following one
\begin{equation}
2g\!\sum_{I\not = J \not = K}{\rm
Tr}\Big(C^{\mu\nu}\,\overline\varphi^{IJ}D_\mu\overline\varphi^{JK}D_\nu\overline\varphi^{KI}
\Big)
\label{antiholsuppot}
\end{equation}
whose origin can be simply traced in the $C^{\alpha\beta}$--term
of the $\star$-product definition (see
Eq. (\ref{starproduct})).

In the following we will show that all structures of the NAC quiver gauge theory
we have presented here are reproduced in a natural and efficient way by the
dynamics of fractional D-branes in a graviphoton R-R background.

\section{${\cal N}=1$ gauge theories from open strings in
$\mathbb{C}^3/(\mathbb{Z}_2\times\mathbb{Z}_2)$}
\label{sec:frac}
It is well-known that quiver gauge theories \cite{quivers} such as the one
considered in section \ref{sec:gauge} may be derived from a
consistent string theory construction; in fact they describe the dynamics of
massless modes of the open strings attached to systems of fractional branes
in a space whose ``internal'' directions are orbifolded by some discrete group.
The type of orbifold one takes determines the amount of residual
supersymmetry and the shape of the quiver.
Indeed, in the stringy interpretation, the nodes of the
quiver correspond to the various types of fractional branes, which in turn
correspond to the irreducible representations of the orbifold group
\cite{fractional}.
To engineer a gauge theory in four dimensions we will consider stacks of parallel
D3 branes, and mod out the six-dimensional transverse space
by the action of a discrete $\mathrm{SU}(3)$ subgroup
in order to remain with four real supercharges, {\it i.e.} with $\mathcal{N}=1$
supersymmetry; in particular we will consider the
$\mathbb{C}^3/(\mathbb{Z}_2\times\mathbb{Z}_2)$ orbifold
which yields exactly the quiver of Fig. \ref{fig:quiver}.

\subsection{The $\mathbb{C}^3/(\mathbb{Z}_2\times\mathbb{Z}_2)$ orbifold and its
conformal fields}
\label{subsec:orbCFT}
Let us consider Type IIB string theory in
$\mathbb{R}^{4}\times \mathbb{C}^3/(\mathbb{Z}_2\times\mathbb{Z}_2)$. To define the
orbifold, we first complexify the ``internal'' coordinates
$x^a \equiv x^5,\ldots,x^{10}$ and the corresponding
string fields $X^a$ and $\psi^a$ by setting
\begin{equation}
\label{frac1}
\begin{aligned}
Z^1 &= (X^5 +\ii X^6)/\sqrt{2}~~~,~~~    & \Psi^1 &= (\psi^5 + \ii \psi^6)/\sqrt{2}~~,
\\Z^2 &= (X^7 +\ii X^8)/\sqrt{2}~~~,~~~    & \Psi^2 &= (\psi^7 + \ii
\psi^8)/\sqrt{2}~~,
\\Z^3 &= (X^9 +\ii X^{10})/\sqrt{2}~~~,~~~ & \Psi^3 &= (\psi^9 +
\ii \psi^{10})/\sqrt{2}~~.
\end{aligned}
\end{equation}
Then, we mod out the action of a
$\mathbb{Z}_2\times\mathbb{Z}_2\subset\mathrm{SO}(6)$ group generated by
\begin{equation}
\label{frac2bis}
g_1 = \ee^{\ii\pi (J_2 - J_3)}~~~~,~~~~
g_2 = \ee^{\ii\pi (J_1 - J_3)}
\end{equation}
where $J_{1,2,3}$ are the generators of rotations in the 5-6, 7-8 and
9-10 planes respectively.
Explicitly, we have
\begin{equation}
\label{frac2}
\begin{aligned}
g_1:~(Z^1,Z^2,Z^3) &\rightarrow (Z^1,-Z^2,-Z^3)~~,\\
g_2:~(Z^1,Z^2,Z^3) &\rightarrow (-Z^1,Z^2,-Z^3)~~,
\end{aligned}
\end{equation}
and similarly for $\Psi^{1,2,3}$.

We may summarize the transformation properties (\ref{frac2}) for the conformal
fields $\partial Z^i$ and $\Psi^i$ ($i=1,2,3$) in the Neveu-Schwarz sector by
means of the following table:
\begin{equation}
\label{ZPsirep}
\begin{tabular}{c|c}
conf. field & irrep \\
\hline
$\Big.\partial Z^i$, $\Psi^i$ & $R_i$
\end{tabular}~,
\end{equation}
where $\{R_I\}=\{R_0,R_i\}$ are the irreducible representations of $\mathbb{Z}_2\times
\mathbb{Z}_2$, identified by writing the character table of the group
%as follows ($e$ stands for the identity element):
\begin{equation}
\label{frac3}
\begin{tabular}{c|cccc}
 & $e$ & $g_1$ & $g_2$ & $g_1 g_2$ \\
\hline
$R_0$ & 1 & ~1 & ~1 & ~1 \\
$R_1$ & 1 & ~1 & $-1$ & $-1$ \\
$R_2$ & 1 & $-1$ & ~1 & $-1$ \\
$R_3$ & 1 & $-1$ & $-1$ & ~1
\end{tabular}
\end{equation}
%All irreps are one-dimensional, since the group is abelian.
The Clebsh-Gordan series for these representations is simply given by
\begin{equation}
\label{frac4}
R_0\otimes R_I = R_I~~~~,~~~~
R_i\otimes R_j = \delta_{ij}  R_0 + |\epsilon_{ijk}| R_k~~,
\end{equation}
and will be crucial in determining the open string spectrum.

To analyze the Ramond sector, we must consider the action of the orbifold group
on  spin fields and spinor states. The spin fields are
best described within the bosonized version of the $\mathrm{so}(6)$ current
algebra generated by the world-sheet fermions.
So we set
\begin{equation}
\label{frac5}
\Psi^i = c_i~\ee^{\pm \ii\,\varphi_i}~~~~,
~~~~\overline\Psi^i = c_i~\ee^{- \ii\,\varphi_i}
\end{equation}
where $c_i$ are cocycle factors needed to maintain the fermionic statistic, and
$\varphi_i$ are free bosons with propagators $\langle\varphi_i(z)\varphi_j(z)\rangle
= -\delta_{ij}\log(z-w)$. The currents corresponding to the Cartan generators
are
\begin{equation}
\label{cartans}J_i= -\ii \normord{\psi^{3+2i}\psi^{4+2i}} \,\,=\,\,
\normord{\Psi^i\overline\Psi^i}\,\,= \ii\,\partial\varphi_i~~,
\end{equation}
while the spin fields $S^A \sim \ee^{\ii
\vec\lambda^A\cdot\vec\varphi}$ are associated to the
$\mathrm{so}(6)$ spinor weights
%\footnote{The spinor is chiral (resp, anti-chiral) if the product of the signs
%in \eq{spinorweights} is positive (resp. negative).}
\begin{equation}
\label{spinorweights}
\vec\lambda^A = \frac 12(\pm,\pm,\pm)~,\hskip 0.8cm (A=1,\ldots,8)~.
\end{equation}
Using this information, we easily deduce from (\ref{frac2bis}) the
transformation properties of the various spin fields under the orbifold
generators which are summarized in the following table
\begin{equation}
\label{spintransf}\begin{tabular}{c|c|c|c|c}
anti-chiral &  chiral  & $g_1$ & $g_2$ & irrep \\
\hline
$\Big.S^{---}\equiv\ee^{-\frac {\ii}{2}(\varphi_1 + \varphi_2 + \varphi_3)}$ &
$S^{+++}\equiv\ee^{\frac {\ii}{2}(\varphi_1 + \varphi_2 + \varphi_3)}$ &
$~1$   & $~1$   & $R_0$ \\
$\Big.S^{-++}\equiv\ee^{\frac {\ii}{2}(-\varphi_1 + \varphi_2 + \varphi_3)}$ &
$S^{+--}\equiv\ee^{\frac {\ii}{2}(\varphi_1 - \varphi_2 - \varphi_3)}$ &
$~1$   & $-1$  & $R_1$ \\
$\Big.S^{+-+}\equiv\ee^{\frac {\ii}{2}(\varphi_1 - \varphi_2 + \varphi_3)}$ &
$S^{-+-}\equiv\ee^{\frac {\ii}{2}(-\varphi_1 + \varphi_2 - \varphi_3)}$ &
$-1$  & $~1$   & $R_2$ \\
$\Big.S^{++-}\equiv\ee^{\frac {\ii}{2}(\varphi_1 + \varphi_2 - \varphi_3)}$ &
$S^{--+}\equiv\ee^{\frac {\ii}{2}(-\varphi_1 - \varphi_2 + \varphi_3)}$ &
$-1$  & $-1$  & $R_3$
\end{tabular}
\end{equation}

In the sequel, we will need also the transformation properties of the conformal
operators corresponding to the roots of $\mathrm{so}(6)$, which will play the role
of \emph{auxiliary} fields for various $\mathcal{N}=1$ multiplets in the field theory.
Recalling that the twelve root vectors of $\mathrm{so}(6)$ are $(0,\pm,\pm)$,
$(\pm,0,\pm)$, $(\pm,\pm,0)$, from (\ref{frac2bis}) we
find
\begin{equation}
\label{auxrep}
\begin{tabular}{c|c|c|c}
current & $g_1$ & $g_2$ & irrep \\
\hline
$\Big.\ee^{\ii(\pm\varphi_2\pm\varphi_3)}$ & $~1$   & $-1$  & $R_1$ \\
$\Big.\ee^{\ii(\pm\varphi_1\pm\varphi_3)}$ & $-1$   & $~1$  & $R_2$ \\
$\Big.\ee^{\ii(\pm\varphi_1\pm\varphi_2)}$ & $-1$   & $-1$  & $R_3$
\end{tabular}
\end{equation}
Notice that these twelve currents correspond to operators of the
form $\Psi^i\Psi^j$, $\Psi^i\overline\Psi^j$ and their complex conjugate with $i\not = j$.

The transformation properties under the orbifold group of the various conformal
fields determine which states of the string spectrum survive the projection. For
open strings, however, one has to take into account also the behaviour of the
boundary conditions under the orbifold group. The irreducible consistent
boundary conditions for open strings are known as fractional branes and are
classified by the irreducible representations of the orbifold group. This means
that the endpoint of an open string attached to a fractional brane of type $I$
transforms in the representation $R_I$. Therefore, to determine which states in
the spectrum of an open string stretching between branes of type $I$ and $J$ are
invariant, it is necessary to look for trivial factors in the decomposition of
$R_I\otimes R_J\otimes Q$, where $Q$ is the representation acting on the string
fields, as indicated in the tables (\ref{ZPsirep}), (\ref{spintransf}) and
(\ref{auxrep}). Hence, given the Chan-Paton representations $R_I$ and $R_J$ for
the endpoint, the conformal fields creating an invariant state must transform
only in certain representations, and all this information is efficiently encoded
in a quiver diagram, like the one in Fig. \ref{fig:quiver}.

\subsection{The gauge multiplets}
\label{subsec:gaugemult}
Let us consider a string attached with both ends to branes of the same
type, say $I$. This means that its endpoints do not transform under the orbifold group,
since $R_I\otimes R_I = R_0$ as one can see from (\ref{frac4}). Therefore also the oscillator
part of any surviving state must be invariant under the orbifold.
For example, in the NS sector the states $\psi^\mu_{-\frac 12}\ket 0$ survive, but
none of the states $\Psi^i_{-\frac
12}\ket 0$ does.

More generally, given a stack of $N_I$ branes of type $I$, the
massless open string excitations organize in a $\mathcal{N}=1$ vector multiplet
for the group $\mathrm{U}(N_I)$ produced by the following vertex operators
\begin{equation}
\label{gaugemultNS}
V_A(p) = A_\mu(p)\,\frac{\psi^\mu}{\sqrt
2}\,{\rm e}^{-\phi}\,{\rm e}^{\ii p\cdot
X}
\end{equation}
in the NS sector, and~\footnote{Comparing with Ref.~\cite{Billo:2004zq},
we have included a factor of $\ii$ in the vertex of the gluino $\lambda$ in order to be
consistent with the notation of section \ref{sec:gauge}.}
\begin{equation}
\label{gaugemultR}
\begin{aligned}
V_\lambda(p) &=\,\ii\,
\lambda^{\alpha}(p)\,S_\alpha\,S^{---}\,
{\rm e}^{-\frac{1}{2}\phi}\,{\rm e}^{\ii p\cdot X}~~,
\\
V_{\overline\lambda}(p) &=
\overline\lambda_{\dot\alpha}(p)\,S^{\dot\alpha}
\,S^{+++}\,{\rm e}^{-\frac{1}{2}\phi}\,
{\rm e}^{\ii p\cdot X}
\end{aligned}
\end{equation}
in the R sector, with $S_\alpha$ and $S^{\dot\alpha}$ being the chiral and
anti-chiral spin fields along the world-volume directions.
The polarizations $A_\mu$, $\lambda^{\alpha}$ and $\overline\lambda_{\dot\alpha}$
carry Chan-Paton indices in the adjoint representation of $\mathrm{U}(N_I)$.
In the following, we will adopt the same notation of section
\ref{sec:gauge}, and use $A_\mu^{I}$, $\lambda^I_{\alpha}$
and $\overline\lambda^I_{\dot\alpha}$
to denote the gauge multiplet living on fractional branes of type $I$.
In writing the vertex operators (\ref{gaugemultNS}) and (\ref{gaugemultR})
 we have set $2\pi\alpha'=1$,
and we will consistently do so henceforth.
Appropriate powers of $2\pi\alpha'$ can be easily reinstated in our formulas so as to give
$A_\mu$ dimensions of (length)$^{-1}$, and to the gauginos   $\lambda_{\alpha}$
and $\overline\lambda_{\dot\alpha}$ dimensions of (length)$^{-3/2}$.

As a matter of fact, also the {auxiliary} field $D$ of the
$\mathcal{N}=1$ vector multiplet admits a stringy realization. In
fact, it can be associated to the following
(non-BRST invariant) vertex in the 0-superghost picture of the NS
sector \cite{FID_string}
\begin{equation}
\label{Dvertex}
V_D(p) =
\frac{1}{3}\,D(p)\, \delta_{ij}
:\Psi^i\,\overline\Psi^j:\,{\rm e}^{i p \cdot X}
= \frac{2\ii}{3}\,D(p) \Big(\sum_i\partial\varphi_i\Big)
\,{\rm e}^{\ii  p \cdot X}~~.
\end{equation}

The vertices we introduced above are connected with each other
through the action of the supersymmetry charges
\begin{equation}
Q_\alpha = \oint \frac{dz}{2\pi\ii}\,j_\alpha(z)~~~~{\rm and}~~~~
\overline Q_{\dot\alpha} = \oint \frac{dz}{2\pi\ii}\,j_{\dot\alpha}(z)
\label{susycharges}
\end{equation}
where the currents (in the $(-\frac{1}{2})$-picture) are given by
\begin{equation}
\label{orbisusy}
j_{\alpha}(z) = S_\alpha(z)\, S^{---}(z)\,\ee^{-\frac{1}{2}\phi(z)}~~~~,
~~~~
\overline j_{\dot\alpha}(z)
=  S_{\dot\alpha}(z)\, S^{+++}(z)\,\ee^{-\frac{1}{2}\phi(z)}~~.
\end{equation}
For instance, it is easy to see that
\begin{equation}
\label{Dauxsusy}
\begin{aligned}
\comm{\xi Q}{V_D(w;p)} &=
\xi^\alpha \oint_w \frac{dz}{2\pi\ii}\, j_\alpha(z) \,
V_D(w;p)\\
&= -\,\xi^\alpha\, D(p)\, S_\alpha(w)\,S^{---}(w)\,
\ee^{-\frac{1}{2}\phi(w)}\,{\rm e}^{i p \cdot X(w)}~~.
\end{aligned}
\end{equation}
Upon comparison with (\ref{gaugemultR}), we recognize in the the
last line the vertex operator of a gaugino, and thus we can rewrite
(\ref{Dauxsusy}) as
\begin{equation}
\label{Dauxsusy2}
\comm{\xi Q}{V_D(w;p)} = V_{\delta \lambda}(w;p)
\end{equation}
with $\delta\lambda = \ii\,\xi D$, in agreement with the standard definitions
in supersymmetric field theory (see Eq. (\ref{susy})).
With similar calculations one can reconstruct
also the other terms in the supersymmetry transformations of the
$\mathcal{N}=1$ gauge multiplet.

\vskip 0.5cm
\subsection{The chiral multiplets}
\label{subsec:chiralmult}
The massless spectrum of open strings stretching between a fractional brane of
type $I$ and one of type $J$ is produced by vertex operators which
transform in some non-trivial representation of the orbifold group, as indicated by the
quiver diagram in Fig. \ref{fig:quiver}. Let us consider, for example,
the oriented open strings stretching between branes of type 0 and type 1.
Then, from (\ref{frac4}) we see that the vertices surviving the orbifold projection
must transform in the representation $R_1$. At the massless level we find
\begin{equation}
\label{chirverticesNS}
\begin{aligned}
V_{\varphi^{01}}(p) &=
\frac{g}{2}
\,\varphi^{01}(p)\,\overline\Psi^1\,\ee^{-\phi}\,\ee^{\ii
p\cdot X}~~,
\\
V_{F^{01}}(p) &=
g \,F^{01}(p)\,\Psi^2\Psi^3\,{\rm e}^{\ii p\cdot
X}
\end{aligned}
\end{equation}
in the NS sector, and
\begin{equation}
V_{\chi^{01}}(p) =
\frac{g}{\sqrt{2}}
\,\chi^{01\,\alpha}(p)\,S_\alpha\,S^{-++}\,{\rm
e}^{-\frac{1}{2}\phi}\,{\rm e}^{\ii p\cdot X}
\label{chirverticesR}
\end{equation}
in the R sector. These are precisely the vertices for the fields (including the
auxiliary one) of a chiral
supermultiplet which, following the notation of section
\ref{sec:gauge}, we organize in the superfield $\Phi^{01}$.
The polarizations in (\ref{chirverticesNS}) and
(\ref{chirverticesR}) carry a superscript $^{01}$ which specifies the boundary
conditions of the open strings under consideration, and the normalizations
of the vertex operators are fixed in order to obtain the canonical
action in the field theory
limit, as we shall see later.

Target-space supersymmetry connects the above vertices among each
other in the standard way. Indeed, in analogy with
(\ref{Dauxsusy}), one can show, for example, that
\begin{equation}
\label{susyF}
\comm{\xi Q}{V_{F^{01}}(p)} = V_{\delta\chi^{01}}(p)~~,
\end{equation}
where $\delta \chi^{01} = \sqrt{2}\,\xi F^{01}$
which exactly agrees with (\ref{susy}).

By construction, the chiral superfield $\Phi^{01}$ transforms in the
bifundamental representation $(N_0,\overline N_1)$ with respect to the
$\mathrm{U}(N_0)\times\mathrm{U}(N_1)$ gauge groups defined on the fractional
branes of type $0$ and $1$ respectively~\footnote{Explicitly,
$\Phi^{01}$ is a $N_0\times N_1$ complex matrix
$\left(\Phi^{01}\right)^{i_0}_{~i_1}$ where $i_0=1,\ldots,N_0$ and $i_1=1,\ldots,N_1$. In
the sequel we will adopt a matrix notation without explicit use of indices,
so we'll have to care about the ordering. For example the covariant derivatives
are $D_\mu\varphi^{01}= \partial_\mu \varphi^{01} +\ii A_\mu^{0} \varphi^{01}- \ii
\varphi^{01}A_\mu^{1}$.}.
Its complex conjugate is denoted as $\overline\Phi^{10}$ and
corresponds to open strings oriented from branes of type 1 to branes of type
0 and transforming in the $(\overline N_0,N_1)$ representation.
Notice that there exists also another \emph{independent} chiral
multiplet arising from the strings oriented from branes of type 1 to
branes of type 0, and transforming in the $(\overline N_0,N_1)$
representation. This other multiplet is denoted as
$\Phi^{10}$ and the vertex operators for its component fields have
the same form as those in (\ref{chirverticesNS}) and (\ref{chirverticesR}),
since they must obey again the requirement of belonging to the
representation $R_1$ of the orbifold group.
Altogether, from a generic system of fractional branes in the orbifold
$\mathbb{C}^3/(\mathbb{Z}_2\times\mathbb{Z}_2)$, we find
twelve chiral multiplets $\Phi^{IJ}$ (with $I\not= J = 0,\ldots,
3$) and their complex conjugates, that are associated to the various oriented links
of the quiver diagram in Fig. \ref{fig:quiver}.

\vskip 0.5cm
\subsection{Effective Lagrangians}
\label{subsec:effectiveactions}
The effective Lagrangians for the massless multiplets of the
quiver theory can be derived by taking the field theory limit $\alpha' \to 0$
of string scattering amplitudes
involving the vertex operators introduced before.
For example, for the gauge multiplet of type $I$, we
must consider diagrams with the vertex operators
(\ref{gaugemultNS}),
(\ref{gaugemultR}) and (\ref{Dvertex})
emitted from disks whose boundaries are entirely attached to branes of
type $I$. In the field theory limit these amplitudes lead to~\footnote{Various details on these
calculations can be found, e.g., in Ref.~\cite{Billo:2004zq}.}
\begin{equation}
\mathcal{L}_{\mathrm{gauge}} =
\frac{1}{g^2}\,{\rm Tr}
\Big\{\frac{1}{2}\,(F^I_{\mu\nu})^2 -2\ii\,\bar\lambda^I\bar\sigma^\mu
D_\mu\lambda^I - (D^I)^2 \Big\}~~.
\label{n1}
\end{equation}
By introducing a self-dual antisymmetric auxiliary field $H_{\mu\nu}^I$,
it is possible to reduce the quartic interactions in
${\rm Tr}(F_{\mu\nu}^I)^2$ to cubic ones.
Indeed, the Lagrangian
\begin{equation}
\label{n11}
\begin{aligned}
\mathcal{L}'_{\mathrm{gauge}}
&= \frac{1}{g^2} ~\mathrm{Tr}\Big\{\big(\partial_\mu A_\nu^I-\partial_\nu A_\mu^I\big)
\partial_\mu A_\nu^I \,+\, 2{\rm i}\, \partial_\mu A_\nu^I
\big[A_\mu^I,A_\nu^I\big] - (D^I)^2\\
&~~~~~~~~~~~~~-2\ii\,\bar\lambda^I\bar\sigma^\mu
D_\mu\lambda^I
+ \frac{1}{4}\,(H_{\mu\nu}^I)^2+ H_{\mu\nu}^I\,\big[A_\mu^I,A_\nu^I\big] \Big\}%~~.
\end{aligned}
\end{equation}
is easily seen to be equivalent to (\ref{n1}) after integrating out
$H_{\mu\nu}^I$ through its algebraic equations of motion.
The auxiliary field $H_{\mu\nu}^I$ admits a stringy representation in terms
of the following (non-BRST invariant) vertex operator in the 0-superghost
picture~\cite{Billo:2004zq}
\begin{equation}
V_{H_{\mu\nu}}(p) =
\frac{1}{2}\,H_{\mu\nu}^I(p)\,\normord{\psi^\nu\psi^\mu}\,{\rm
e}^{\ii p \cdot X}~~.
\label{vhmunu}
\end{equation}
Notice that the structure of this vertex is the same as that of the
$p\cdot\psi \psi^\mu$ part of the (properly normalized) vertex for the gauge field in the
0-picture, namely
\begin{equation}
\label{vertA0}
{V}_{A}(p) = 2\ii\,A_\mu^I(p)\left(\partial X^\mu + \ii \,p\cdot \psi
\,\psi^\mu \right)\ee^{\ii p \cdot X}~~.
\end{equation}
Thus, whenever in a disk amplitude we get a non-vanishing amplitude by inserting a
vertex in the 0-picture for $A_\mu^I$, we get also a non-vanishing
amplitude by inserting the vertex (\ref{vhmunu}) for $H_{\mu\nu}^I$.

It is worth pointing out that this auxiliary field is useful not only
to reduce the quartic interactions in the gauge Lagrangian to cubic ones, but also to
linearize the supersymmetry transformations of the vector multiplet.
For example, by using the vertices (\ref{vertA0}) and (\ref{vhmunu}) and computing
\begin{equation}
\comm{\xi Q}{V_{A}(p)}~~~~{\rm and}~~~~\comm{\xi Q}{V_{H_{\mu\nu}}(p)}
\label{commut}
\end{equation}
one easily obtains the following supersymmetry transformation for
the gaugino
\begin{equation}
\delta \lambda = -\xi\,\sigma^{\mu\nu}\big(\partial_\mu A_\nu -
\frac{\ii}{4}\,H_{\mu\nu}\big)
\label{susylambda}
\end{equation}
which, upon eliminating $H_{\mu\nu}$ through its field equation,
becomes $\delta \lambda = -\frac{1}{2}\,\xi\,\sigma^{\mu\nu} F_{\mu\nu}$,
with the non-linear terms of the field strength included (see Eq. (\ref{susy})).

A similar analysis can be done also in the matter sector. By
computing all disk diagrams with insertions of the vertex
operators (\ref{chirverticesNS}) and (\ref{chirverticesR}), and
then taking the field theory limit, one may reconstruct the effective
action for the chiral multiplet $\Phi^{01}$ (and its complex conjugate
$\overline\Phi^{10}$) which is given by
\begin{equation}
\label{kinchir}
\begin{aligned}
{\cal L}_{\rm matt}
\,=&\,\mathrm{Tr}\Big\{D_\mu\overline\varphi^{10}D_\mu\varphi^{01}
-\ii\,\overline\chi^{10}\bar\sigma^\mu D_\mu\chi^{01}
+\overline F^{10}F^{01}\\
&+\overline\varphi^{10}D^0\varphi^{01}-\varphi^{01}D^1\overline\varphi^{10}
+\sqrt 2\,{\ii}\,\big(\overline\chi^{10}\overline\lambda^0\varphi^{01}
-\varphi^{01}\overline\lambda^1\overline\chi^{10}\big)
\\&+\sqrt 2\,{\ii}\,\big(
\overline\varphi^{10}\lambda^0\chi^{01}
-\chi^{01}\lambda^1\overline\varphi^{10}\big)\Big\}~~.
\end{aligned}
\end{equation}
Notice that the disk diagrams which lead to this effective
Lagrangian have their boundaries lying partly on branes of type 0
and partly on branes of type 1, and consequently the gauge fields
can be of either type.

As for the gauge sector, also in the matter part it is
possible to introduce suitable auxiliary fields to
decouple the quartic interactions coming from the covariant derivatives of the scalars.
Indeed, the very first term of \eq{kinchir} can be rewritten as
\begin{equation}
\label{kinchir2}
\begin{aligned}
\Tr\Big\{\partial_\mu \overline\varphi^{10}\partial_\mu \varphi^{01}
 &+\overline H_\mu^{10} H_\mu^{01}+\ii\big(\partial_\mu \overline\varphi^{10} - \ii\,\overline
 H_\mu^{10}\big)
 \big(A_\mu^0\varphi^{01}-\varphi^{01}A_\mu^1\big)\\
 &+\ii\big(A_\mu^1\overline\varphi^{10}-\overline\varphi^{10}A_\mu^0\big)
 \big(\partial_\mu \varphi^{01} - \ii\, H_\mu^{01}\big)\Big\}~~,
\end{aligned}
\end{equation}
{\it i.e.} with only cubic interactions. The new $H$-dependent terms arise from
disk diagrams with insertions of the following  auxiliary vertex operator
\begin{equation}
\label{Hmuvert}
V_{H^{01}_\mu}(p) =  g\,
H^{01}_\mu(p) \,\psi^\mu\,\overline\Psi^1\,\ee^{\ii p\cdot X}
\end{equation}
and of the corresponding complex conjugate, as shown in Fig.
\ref{fig:Hmuint}.
\FIGURE{\centerline{
\psfrag{A0}{$A_\mu^0$}
\psfrag{A1}{$A_\mu^1$}
\psfrag{phib10}{$\overline\varphi^{10}$}
\psfrag{Hmu01}{$H_\mu^{01}$}
\includegraphics[width=0.55\textwidth]{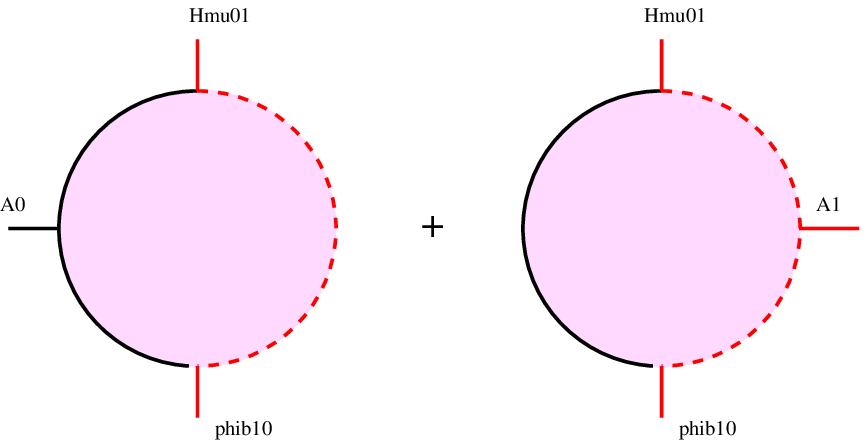}%
\caption{\small The diagrams accounting for the interactions of the auxiliary
field $H^{01}_\mu$.}
\label{fig:Hmuint}
}}

Notice that this vertex is identical to the fermionic part of the
(properly normalized) scalar vertex in the 0-superghost picture,
namely
\begin{equation}
\label{phivertex0pic}
V_{\varphi^{01}}(p) =
\sqrt{2}\ii \,g\,
\varphi^{01}(p)\,
\left(\partial \overline Z^1 + \ii p\cdot\psi\,
\overline\Psi^{1}\right)\,\ee^{\ii p\cdot X}~~.
\end{equation}
Thus everywhere we have to consider a diagram with a
vertex for $\varphi^{01}$ in the 0-picture (which produces terms
containing $\partial_\mu\varphi^{01}$
in the Lagrangian), we have also to consider a diagram with
the auxiliary vertex (\ref{Hmuvert}).
The net outcome of this is that all occurrences
of  $\partial_\mu\varphi^{01}$ in the effective action are
promoted to $\big(\partial_\mu\varphi^{01}
-\ii H^{01}_\mu\big)$, which in turn becomes the complete covariant
 derivative $D_\mu\varphi^{01}$ after integrating out $H_\mu^{01}$ through its
field equation.

Let us now turn to the superpotential. Whenever the string configuration contains
at least three different types of branes, then the Chan-Paton
structure of the vertex operators for chiral multiplets allows
for a cubic holomorphic superpotential. Let us suppose, for example, to have
branes of type $0$, $3$ and $1$. Then, if consider a disk diagram
with three vertices corresponding to some of the fields in the
multiplets $\Phi^{03}$, $\Phi^{31}$ and $\Phi^{10}$, taken in this order,
we have the possibility of getting a non-vanishing result. Indeed, the disk
boundary jumps first from type 0 to type 3, then from 3 to 1 and finally returns
back to type 0 to close in a consistent way. Of course, we could get a non-zero
amplitude also by inserting the vertices corresponding to
$\Phi^{01}$, $\Phi^{13}$ and $\Phi^{30}$, {\it i.e.} by following the triangle
on the quiver diagram in the opposite direction, or by utilizing the anti-chiral
counterparts
of the above possibilities which lead to a cubic anti-holomorphic
superpotential
\footnote{Notice that the Chan-Paton structure allows in principle
configurations that involve both holomorphic and anti-holomorphic
fields, like for example $\Phi^{03}$, $\overline\Phi^{31}$ and
$\Phi^{10}$. However, the corresponding amplitudes vanish since
the vertex operators in these configurations do not saturate the charges with respect
to the internal world-sheet bosons $\varphi_i$.
Thus, only holomorphic or anti-holomorphic superpotentials are possible.}.

Specifically, if we compute the amplitude among $V_{F^{03}}$, $V_{\varphi^{31}}$
and $V_{\varphi^{10}}$ and take the field theory limit, we obtain the
following term in the effective Lagrangian
\begin{equation}
\label{yukawaauxresult}
g\,\Tr \Big(F^{03} \varphi^{31} \varphi^{10}\Big)~~,
\end{equation}
which is related by supersymmetry to the Yukawa term
\begin{equation}
\label{yukawa}
 -\,g\, \Tr\Big(\varphi^{03}\chi^{31}\chi^{10}\Big)
\end{equation}
arising from the amplitude among $V_{\varphi^{03}}$, $V_{\chi^{31}}$ and
$V_{\chi^{10}}$ (see also the first line of Eq. (\ref{holsuppot})).
These interactions as well all others
corresponding to different combinations of fields can be
summarized in a holomorphic superpotential of the form
\begin{equation}
\label{suppot}
W = \frac{g}{3}\,\sum_{I\not= J \not= K} \Tr
\Big(\Phi^{IJ}\Phi^{JK} \Phi^{KI})
\end{equation}
or in its anti-holomorphic counterpart.

\section{NAC deformation from R-R flux}
\label{sec:def}
We now analyze the deformations of the $\mathcal{N}=1$ quiver theory discussed
in the previous section that are induced by a non-trivial R-R flux corresponding
to a graviphoton background with constant field strength. This background is
described by a constant antisymmetric tensor $C_{\mu\nu}$ which we take to be
self-dual and which is responsible of the NAC deformation of the $\mathcal{N}=1$
superspace. From the string point of view, $C_{\mu\nu}$ is a R-R field strength,
and more precisely it is the R-R 5-form of Type II B string theory%
\footnote{The effect of a constant RR 5-form field-strength background
in the $\mathcal{N}=4$ case has been
recently considered in \cite{Imaanpur:2005pd}.}, wrapped
around the internal orbifold space and described by the following closed string
vertex operator (in the $(-1/2,-1/2)$ superghost picture)
\begin{equation}
\label{RRvop}
V_{C}(z,\overline z)
=\frac{1}{4\pi^2}
\,C^{\alpha\beta}\,S_{\alpha}(z)\,S^{---}(z)\,
\ee^{-\frac{1}{2}\phi(z)}\,\, {\widetilde S}_{\beta}(\overline z)
\,{\widetilde{S}}^{---}(\overline z)\,\ee^{-\frac{1}{2}\tilde\phi(\overline
z)}~~.
\end{equation}
Here, using the arguments explained in Ref.
\cite{Billo:2004zq}, we have already identified the symmetric bispinor
polarization of $V_{C}$ with the non-anti-commutativity
parameter $C^{\alpha\beta}$ used in section \ref{sec:gauge}.
It is worth recalling that such parameter has dimensions of
(length), and thus a factor of $(2\pi\alpha')^{-1/2}$ should be
included in right hand side of (\ref{RRvop}) to make $V_{C}$
adimensional. Even if we are using conventions in which $2\pi\alpha'=1$,
these dimensional considerations will be crucial in the following.
In (\ref{RRvop}) the tilde
denotes the right movers of the closed string, and $z$ a point in the upper-half complex
plane, which is conformally equivalent to the interior of a disk.
Notice that the vertex operator $V_C$ does not contain the usual
plane wave term $\ee^{\ii p\cdot X}$, since we are considering a
{\it constant} background with $p=0$.

We are now going to systematically study string amplitudes for
fractional D3 branes of the orbifold $\mathbb{C}^3/(\mathbb{Z}_2 \times
\mathbb{Z}_2)$ in the presence of the non-trivial R-R background
(\ref{RRvop}), {\it i.e.} we shall compute mixed open/closed
string amplitudes on disks with different types boundary
conditions corresponding to the various types of branes.
In any mixed open/closed string amplitudes on a disk,
the presence of the boundary forces an identification
between left- and right-moving oscillators
of the closed strings. In our specific case, this identification is the same
on all types of fractional D3 branes, and amounts, in practice, to
the following replacements (see e.g. Ref. \cite{Billo:2004zq} for details)
\begin{equation}
\label{ident}
{\widetilde S}_{\alpha}(\overline z) \to S_{\alpha}(\overline z)~~~~,~~~~
{\widetilde S}^{---}(\overline z) \to {\widetilde S}^{---}(\overline
z)~~~~,~~~~
\widetilde\phi(\overline z) \to \phi(\overline z)~~.
\end{equation}
As a consequence, we see that any insertion of the R-R vertex (\ref{RRvop})
carries an effective charge $(-1,-1,-1)$ with respect to the three
world-sheet scalars $\varphi_1$, $\varphi_2$ and $\varphi_3$ which
bosonize the fermions in the internal orbifold directions, and an
effective superghost charge $(-1)$.
Therefore, in order to have a non vanishing amplitude with a
single R-R insertion, we have to choose the open string vertex
operators in such a way that they carry an effective total charge
$(+1,+1,+1)$ with respect to the internal world-sheet bosons, and
an effective total superghost charge $(-1)$. These new
requirements add to the one of having a consistent Chan-Paton
structure, which we already encountered.
Furthermore, we are interested only in amplitudes which survive in
the field theory limit $\alpha'\to 0$ with $g$ fixed. Since the factors of
$(2\pi\alpha')^h$ (which we have not written explicitly) in the definitions
of the vertex operators  give space-time dimensions of (length)$^{-h}$ to the corresponding
polarizations, this implies that we have to look only for structures with total dimension
of (length)$^{-4}$, including the contribution of the R-R field which carries dimension
of (length).

Let us now analyze the effects produced by the insertion of the R-R vertex
(\ref{RRvop}) in the various sectors.

\subsection{The gauge and chiral matter sectors}
\label{subsec:gauge_sector}
The study of the R-R deformation in the pure gauge sector has been
the subject of Ref. \cite{Billo:2004zq}, where it was shown that
the only string amplitudes on disks with a single type of
boundary that do not vanish in the field theory limit, are
\begin{equation}
\label{gaugeRRampli}
\lvev
V_{\overline\lambda^{I}}\,V_{\overline\lambda^{I}}\,V_{A_\mu^{I}}\,
V_C \rvev~~~~\text{and}~~~~
\lvev
V_{\overline\lambda^{I}}\,V_{\overline\lambda^{I}}\,V_{H_{\mu\nu}^{I}}\,
V_C \rvev~~,
\end{equation}
and correspond to the following
contribution to the effective Lagrangian for the $\mathrm{U}(N_I)$ gauge
fields \footnote{In Ref. \cite{Billo:2004zq} we used a $C^{\mu\nu}$ that
is twice the one used in the present paper and in the majority of the
literature. We have taken this difference into account in writing
\eq{defgaugeac}.}
\begin{equation}
\label{defgaugeac}
\frac{4\ii}{g^2}\,C^{\mu\nu}\,\Tr\Big\{
 \big(\partial_\mu A_\nu^{I}
 - \frac{\ii}{4} \,H_{\mu\nu}^{I}\big)
\overline\lambda^{I}\overline\lambda^{I}\Big\}~~.
\end{equation}
When this term is added to the undeformed Lagrangian (\ref{n11})
and the auxiliary field $H_{\mu\nu}^I$ is integrated out, one
recovers precisely the $\mathcal{N}=1/2$ gauge Lagrangian (\ref{ymlagrangian'})
that follows from the NAC deformation of the superspace.

In the case of disks with more than one type of boundary, we have more
possibilities. For example a consistent Chan-Paton structure and a correct
balance of internal charges can be obtained by inserting $V_C$ together with a
vertex for $\overline\varphi^{10}$ (with charges $(1,0,0)$), and a vertex for
$F^{01}$ (with charges $(0,1,1)$). As far as the superghost background charge is
concerned, the vertex for $V_{F^{01}}$ is defined in the 0-picture, and so we
have to insert the vertex $V_{\overline\varphi^{10}}$ in the $(-1)$-picture to
soak up the superghost number anomaly. However, an amplitude with just
$C^{\alpha\beta}$, $\overline\varphi^{10}$ and $F^{01}$ cannot survive in the
field theory limit, since these fields combined have dimensions of
(length)$^{-2}$ and no momentum factor is produced given the above picture
assignments. The way out, in this example, is clearly to insert a further vertex
in the 0-picture, that carries zero internal charge and supplies the needed mass
dimensions. Such vertices may only come from the gauge multiplets of
$\mathrm{U}(N_0)$ or $\mathrm{U}(N_1)$, and in principle can be either vertices
for the gauge field $A_\mu$, for the $D$ field or for the ``auxiliary'' fields
$H_{\mu\nu}$, which we can insert either on the boundary of type 0 or on the
boundary of type 1.

If we insert a vertex for a $D$ field, {\it i.e.} if we compute
$\lvev V_{\overline\varphi^{10}} V_{D^{0}}V_{F^{01}} V_C \rvev$
or $\lvev V_{\overline\varphi^{10}}V_{F^{01}}V_{D^{1}}
V_C \rvev$, we find that the resulting world-sheet
correlator in the $\mathrm{SO}(4)$ current algebra sector
is
\begin{equation}
\label{epsicorr}
\langle S_{\alpha}(z)\, S_{\beta}(\overline z)
\rangle ~\propto ~\varepsilon_{\alpha\beta}
\end{equation}
which vanishes when it is contracted with the symmetric
polarization $C^{\alpha\beta}$. Thus, we have to consider the other two
possibilities, which are represented in Fig. \ref{fig:kindef}(a)
and \ref{fig:kindef}(b).
\FIGURE{\centerline{
\psfrag{F01}{$F^{01}$}
\psfrag{chi01}{$\chi^{01}$}
\psfrag{Amu0}{$A_\mu^0$}
\psfrag{phib10}{$\overline\varphi^{10}$}
\psfrag{lb0}{$\overline\lambda^0$}
\psfrag{Hmunu0}{$H_{\mu\nu}^0$}
\psfrag{Cab}{$C_{\alpha\beta}$}
\includegraphics[width=0.8\textwidth]{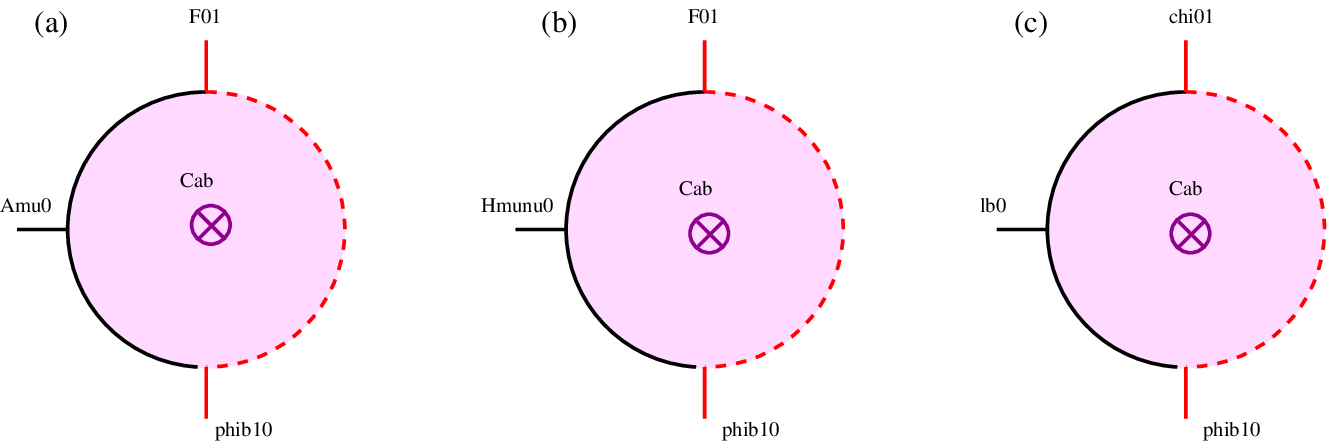}%
\caption{Examples of diagrams with a R-R insertion on a disks with two distinct
types of boundaries.}
\label{fig:kindef}
}}

When we insert the gauge field vertex (\ref{vertA0}), the part containing $\partial X^\mu$
does not contribute, as it leads again to the correlator  (\ref{epsicorr}), while the
fermionic part containing $p\cdot\psi\,\psi^\mu$ produces a non-vanishing
result. We have
\begin{equation}
\label{phiFARR}
\begin{aligned}
\lvev V_{\overline\varphi^{10}} V_{A^{0}}V_{F^{01}} V_C \rvev
\,\,\equiv\,\,C_{\rm disk}&\!\int\frac{\prod_i dy_i\, dz d\overline z}{dV_{\rm
CKG}}\,\,\times \\
&~~~~~\langle V_{\overline\varphi^{10}}(p_1;y_1)\, V_{A^{0}}(p_2;y_2)
\,V_{F^{01}}(p_3;y_3)\,V_C (z,\overline z)\rangle
\end{aligned}
\end{equation}
where $C_{\rm disk} = {4}/{g^2}$
is the normalization of any disk amplitude in our present
conventions (see {\it e.g.} Ref. \cite{Billo:2002hm} for further
details) and $dV_{\rm CKG}$ is the $\mathrm{Sl}(2,\mathbb{R})$
invariant volume element. The insertion points $y_i$ of the open
string vertices are integrated on the real axis with $y_1\geq
y_2\geq y_3$, while the closed string insertion $z$ is integrated
in the upper half complex plane. More explicitly, the amplitude
(\ref{phiFARR}) is
\begin{equation}
\label{phiFARR2}
\begin{aligned}
\frac{\ii}{\pi^2}\,{\rm Tr}\,\Big[ C^{\alpha\beta}\,&
\overline\varphi^{10}(p_1)\,\big(\ii\,p_2^\mu\,A^{0}_\nu(p_2)\big)
\,F^{01}(p_3) \Big]\int\frac{\prod_i dy_i\, dz d\overline z}{dV_{\rm
CKG}}\,\times
\\
& \Big\{\langle\ee^{-\phi(y_1)}\,\ee^{-\frac{1}{2}\phi(z)}\,
\ee^{-\frac{1}{2}\phi(\overline z)}\rangle\,
\langle\Psi^1(y_1)\,\Psi^2\Psi^3(y_3)
\,S^{---}(z) \,S^{---}(\overline z)\rangle
\\
& ~~\langle :\psi^\mu\psi^\nu:(y_2) \,S_{\alpha}(z)\, S_{\beta}(\overline
z)\rangle \,\langle \ee^{\ii p_1\cdot X(y_1)}
\,\ee^{\ii p_2\cdot X(y_2)}\,\ee^{\ii p_3\cdot X(y_3)}
\rangle\Big\}~~.
\end{aligned}
\end{equation}
Using the correlation functions given in appendix \ref{subsec:correlators}, and exploiting
the $\mathrm{SL}(2,\mathbb{R})$ invariance to
fix $y_1\to \infty$, $z\to \ii$ and $\overline z \to-\ii$, we
are left with an integral over the remaining positions $y_2$ and $y_3$, which reads
\begin{equation}
\int_{-\infty}^{+\infty}dy_2\,\int_{-\infty}^{y_2}dy_3\,\frac{1}
{\left(y_2^2+1\right)\left(y_3^2+1\right)} =
\frac{\pi^2}{2}~~.
\label{integral}
\end{equation}
Putting everything together, we finally obtain the
following contribution to the effective Lagrangian
\begin{equation}
\label{phiFARR3}
2\ii \,  \Tr
\Big(C^{\mu\nu}\,\partial_\mu A_\nu^{0}\,F^{01}\overline\varphi^{10}
\Big)~~.
\end{equation}
Instead of the vertex $V_{A^0}$, we could have placed a vertex for $A_\mu^{1}$ on
the boundary portion of type 1, obtaining, a part from the different
ordering of the Chan Paton factors and a different sign, the same result as in \eq{phiFARR3}.
Moreover, as we already observed, we may obtain a non vanishing amplitude also by replacing
the 0-picture vertex for the gauge field with the one for the auxiliary field
$H_{\mu\nu}$. This computation of course generalizes to any disk with
two types of boundary, and so altogether we get the following
contribution to the effective Lagrangian
\begin{equation}
\label{phiFARR4}
2\ii \,
\sum_{J\not= I}\Tr \Big\{C^{\mu\nu}\,\big(\partial_\mu
A_\nu^{I}-\frac{\ii}{4}\,H_{\mu\nu}^{I}\big)
\big(F^{IJ}\overline\varphi^{JI} -
\overline\varphi^{IJ} F^{JI} \big)
\Big\}~~.
\end{equation}
Notice that this term has the same structure as the $C$-dependent term
(\ref{defgaugeac}) that was already present in the pure gauge
sector. Since there are no other diagrams involving the R-R background and
the auxiliary field $H_{\mu\nu}^I$, when we add the two contributions (\ref{defgaugeac})
and (\ref{phiFARR4}) to the undeformed action (\ref{n11}), we find
that the auxiliary field can be eliminated through the following
equation
\begin{equation}
\label{eqHImunu}
H^{I}_{\mu\nu} =
-2 \comm{A^{I}_\mu}{A^{I}_\mu}^{(+)}
- 2\,C_{\mu\nu} \Big(\overline\lambda^{I}\overline\lambda^{I}
+ \frac {g^2}{2} \sum_{J\not= I}
\left(F^{IJ}\overline\varphi^{JI} -
\overline\varphi^{IJ} F^{JI} \right)
\Big)
\end{equation}
where the superscript $^{(+)}$ stands for the self-dual part.
Plugging this identification back in the Lagrangian, and summing over all types
of branes, we find that the deformation terms that must be added the Yang-Mills Lagrangian
of the quiver theory are
\begin{equation}
\label{n2}
\begin{aligned}
\frac{1}{g^2}\,\sum_I {\rm Tr} \Bigg\{
2\ii\, C_{\mu\nu}\,F_{\mu\nu}^I\,
\Big(\overline\lambda^{I}\overline\lambda^{I}
+ \frac {g^2}{2} \sum_{J\not= I}
\left(F^{IJ}\overline\varphi^{JI} -
\overline\varphi^{IJ} F^{JI} \right)
\Big)
\\
- 4\,\det C\,
\Big(\overline\lambda^{I}\overline\lambda^{I}
+ \frac{g^2}{2}  \sum_{J\not= I}
\left(F^{IJ}\overline\varphi^{JI} -
\overline\varphi^{IJ} F^{JI} \right)
\Big)^2
\Bigg\}~~.
\end{aligned}
\end{equation}

So far we have considered disk diagrams with open string
vertices in the NS sector. However, there are non-vanishing
diagrams involving also fermionic vertices from the R sector.
An example of such diagrams is represented in Fig. \ref{fig:kindef}(c)
which corresponds to the amplitude $\lvev V_{\overline\varphi^{10}}
V_{\overline\lambda^{0}} V_{\chi^{01}} V_C\rvev$.
To soak up the superghost charge, we put the vertices
for $\overline\lambda^0$ and $\chi^{01}$ in the $(-1/2)$-picture and
the vertex for $\overline\varphi^{10}$ in the 0-picture as given
in (\ref{phivertex0pic}). Using the explicit expressions for
these vertices and performing the appropriate OPE's, one can
easily compute this string amplitude along the same lines discussed above
and in the end one finds the following contribution to the
effective Lagrangian
\begin{equation}
\label{philambdachiRR}
\sqrt{2}\,C^{\mu\nu}\,{\rm Tr}\Big(\overline\lambda^{0}\bar\sigma_\nu
\chi^{01}\,\partial_\mu\overline\varphi^{10}\Big)
~~.
\end{equation}
Similarly to the case discussed in section \ref{subsec:chiralmult},
besides the previous diagram we have
also to consider the one where the 0-picture vertex $V_{\overline\varphi^{10}}$
is replaced by the vertex for the auxiliary field $\overline H^{10}_\mu$
given in (\ref{Hmuvert}), with the result that
$\partial_\mu\overline\varphi^{10}$ in (\ref{philambdachiRR})
is shifted to $\big(\partial_\mu\overline\varphi^{10} -\ii\, \overline H_\mu^{10}\big)$.
Notice that instead there are no amplitudes involving the R-R background and the
auxiliary field $H_\mu^{01}$, due to unbalanced internal charges.
Therefore, when we add these terms
to the undeformed Lagrangian (\ref{kinchir2}), we find that
the auxiliary field $\overline H^{10}_\mu$ can still be eliminated
through its undeformed equation of motion, namely
\begin{equation}
\label{Hmubareq}
\overline H^{10}_\mu = \overline\varphi^{10} A^{0}_\mu -
A^{1}_\mu \overline\varphi^{10} ~~.
\end{equation}
Again, the net effect is that the ordinary derivative in (\ref{philambdachiRR})
is promoted to the full covariant derivative
$D_\mu\overline\varphi^{10}$ and gauge invariance is restored.
Repeating this analysis for all possible multiplets on various types of boundaries,
we find that the $C$-dependent Lagrangian arising from fermionic
vertices of the R sector is
\begin{equation}
\label{philambdachiRRcov}
\sqrt{2}\,C^{\mu\nu}\,\sum_{J\not= I}\Tr \Big\{
\big(\overline\lambda^{I}\bar\sigma_\nu \chi^{IJ}
-\chi^{IJ}\sigma_\nu\overline\lambda^{J}\big)
D_\mu\overline\varphi^{JI}\Big\}~~.
\end{equation}

Eqs. (\ref{n2}) and (\ref{philambdachiRRcov}) describe the
deformation terms induced by the R-R graviphoton background
(\ref{RRvop}) on the effective action of the quiver gauge theory,
and are strictly related to those that can be obtained using the
NAC $\star$-product deformation described in section \ref{sec:gauge}
(see Eq. (\ref{Lquiver})). To make a simple comparison, let us
concentrate on a single gauge group, say $\mathrm{U}(N_0)$ which
corresponds to the branes of type 0, and on single charged chiral
multiplet, say $\Phi^{01}$ and its conjugate $\overline\Phi^{10}$.
Dropping for ease of notation the indices on such fields, we can
easily see that (\ref{n2}) and (\ref{philambdachiRRcov}) in this case reduce to
\begin{equation}
\label{1gg}
\begin{aligned}
\frac{1}{g^2} {\rm Tr} \Big\{
2\ii\, C^{\mu\nu}F_{\mu\nu}\,&
\Big(\overline\lambda\,\overline\lambda
+ \frac {g^2}{2}\,F\,\overline\varphi \Big)
 - 4\,\det C\,
\Big(\overline\lambda\,\overline\lambda
+ \frac{g^2}{2}\,F\,\overline\varphi\Big)^2
\\
&+ \sqrt{2}\,g^2\,C^{\mu\nu} D_\mu\overline\varphi\,\overline\lambda\,\bar\sigma_\nu
\chi \Big\}~~.
\end{aligned}
\end{equation}
These are precisely the interaction terms that appear in the Lagrangians
(\ref{ymlagrangian'}) and (\ref{mattlagrangian'}) (with
$a'=1$ and $b'=-4$) based on the NAC $\star$-product deformation,
with, in addition, an extra term
\begin{equation}
\label{extra1}
-{g^2}\,\det C\,{\rm Tr}\big(F\,\overline\varphi\big)^2~~.
\end{equation}
This is, however, exactly of the form (\ref{interaction}) (with $c'=-1$).
As we remarked in section \ref{subsec:gaugetheories}, such a term
can be induced by a NAC $\star$-product, provided the auxiliary field
$\overline F$ is shifted according to (\ref{F''}), and is produced
at the 1-loop level.

Finally, it is worth pointing out that if we insert the deformed field equation
(\ref{eqHImunu}) into the linearized supersymmetry transformation
(\ref{susylambda}) for the gaugino $\lambda^I$, we can recover
exactly all non-linear terms of $\delta\lambda^I$ appearing in
(\ref{susy}), including the $C$-dependent ones.

\subsection{The superpotential sector}
\label{subsec:superpotential_sector}
Let us now analyze the effects produced by the insertion of the R-R graviphoton
vertex (\ref{RRvop}) in diagrams with three types of boundary
conditions that contribute to the effective superpotential of the
quiver theory. One specific example is represented in Fig.
\ref{fig:suppotdef1}(a) which describes the following
amplitude $\lvev V_{\overline\varphi^{01}}
V_{\overline\varphi^{13}} V_{\overline\varphi^{30}} V_C\rvev$.
\FIGURE{\centerline{
\psfrag{phib10}{$\overline\varphi^{10}$}
\psfrag{phib03}{$\overline\varphi^{03}$}
\psfrag{phib31}{$\overline\varphi^{31}$}
\psfrag{F10}{$F^{10}$}
\psfrag{F03}{$F^{03}$}
\psfrag{F31}{$F^{31}$}
\psfrag{chi03}{$\chi^{03}$}
\psfrag{chi31}{$\chi^{31}$}
\psfrag{Cab}{\small $C_{\alpha\beta}$}
\psfrag{Cgd}{\small $C_{\gamma\delta}$}
\includegraphics[width=0.9\textwidth]{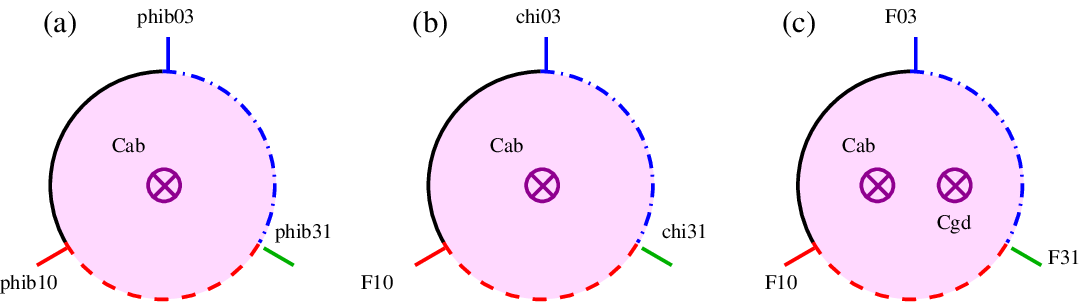}%
\caption{Examples of diagrams with R-R insertions on disks with three distinct
types of boundaries that contribute to the $C$-deformed superpotential.}
\label{fig:suppotdef1}
}}

In order to saturate the superghost charge, one of the three vertices for
the scalars can be put in the $(-1)$-picture and the other two in the $0$-picture.
Computing the corresponding string amplitude, in the field theory
limit we obtain the following contribution to the Lagrangian
\begin{equation}
\label{Cbarphi^3}
2g\,{\rm Tr}\Big(C^{\mu\nu}\,\overline\varphi^{01}
\partial_\mu\overline\varphi^{13}\partial_\nu\overline\varphi^{30}
\Big)~~.
\end{equation}
As in previous cases, also here we should consider the diagram in
which the $0$-picture vertices for $\overline\varphi$ are replaced by the vertices
for the auxiliary fields $\overline H_\mu$, so that in the end the ordinary
derivatives in (\ref{Cbarphi^3}) are promoted to the full
covariant derivatives. Repeating this calculation for all triples
of boundary conditions that can be consistently found in the quiver diagram,
we finally obtain
\begin{equation}
2g\!\sum_{I\not = J \not = K}{\rm
Tr}\Big(C^{\mu\nu}\,\overline\varphi^{IJ}D_\mu\overline\varphi^{JK}D_\nu\overline\varphi^{KI}
\Big)
\label{antiholsuppot1}
\end{equation}
{\it i.e.} precisely one of the $C$-dependent terms of the
anti-holomorphic deformed superpotential (see Eq. (\ref{antiholsuppot})).

Let us now consider the diagram represented in Fig.
\ref{fig:suppotdef1}(b), which corresponds to the amplitude
$\lvev V_{F^{01}} V_{\chi^{13}} V_{\chi^{30}} V_C\rvev$
involving fermionic vertices from the R sector. The evaluation of
this amplitude is strictly analogous to what we have already
described in the previous subsection and, after generalizing to
all triples of consistent boundary conditions, we find
\begin{equation}
\label{holsuppot1}
\frac{g}{4}\!\sum_{I\not = J \not = K}\!{\rm
Tr}\Big(C^{\mu\nu}\,F^{IJ}\chi^{JK}\sigma^{\mu\nu}\chi^{KI}
\Big)
\end{equation}
which is exactly one of the terms expected from the NAC
$\star$-product deformation (see Eq. (\ref{holsuppot})).

Finally, let us analyze the diagram of Fig.
\ref{fig:suppotdef1}(c), which, differently from all other diagrams
considered so far, has two R-R insertions. It corresponds to the
amplitude $\lvev V_{F^{01}} V_{F^{13}} V_{F^{30}} V_C\,V_C\rvev$
which is easily seen to respect all requirements in order to be
non-vanishing and survive in the field theory limit. From the open
string point of view, this is a 3-point amplitude which
cannot be further reduced by means of suitable auxiliary fields,
and thus it has to be evaluated explicitly. Since there are three
open and two closed string insertions, the calculation is more
involved than the ones encountered before, but it is still doable.
More precisely, we have
\begin{equation}
\label{FFFRRRR0}
\begin{aligned}
\lvev V_{F^{01}} V_{F^{13}} V_{F^{30}} &V_C\,V_C\rvev
\,\,\equiv\,\,\frac{1}{2}\,C_{\rm disk}\!\int\frac{\prod_i dy_i\, dz d\overline z
\,dw d\overline w}{dV_{\rm
CKG}}\,\,\times \\
&\langle V_{F^{01}}(p_1;y_1)\,V_{F^{13}}(p_2;y_2)
\,V_{F^{30}}(p_3;y_3)\,V_C (z,\overline z)\,V_C (w,\overline w)\rangle
\end{aligned}
\end{equation}
where the symmetry factor of $1/2$ accounts for the presence of two alike
R-R vertices. Inserting the explicit expressions for the various
ingredients and computing the world-sheet correlators,
the above amplitude becomes
\begin{equation}
\label{FFFRRRR1}
\begin{aligned}
\frac{g}{8\pi^4}\,& {\rm Tr} \Big( C^{\alpha\beta} C^{\gamma\delta}
F^{01}(p_1) F^{13}(p_2)
F^{30}(p_3) \Big)\,
\int\frac{\prod_i dy_i\, dz d\overline z
\,dw d\overline w}{dV_{\rm
CKG}}~\times
\\
& \Big\{
\langle
\ee^{-\frac 12\phi(z)}\,\ee^{-\frac 12\phi(\overline z)}
\,\ee^{-\frac 12\phi(w)}\,\ee^{-\frac 12\phi(\overline w)}
\rangle ~~\langle
S_{\alpha}(z)\, S_{\beta}(\overline z)\,
S_{\gamma}(w)\, S_{\delta}(\overline w)\rangle
\\
&~~
\langle
\Psi^2\Psi^3(y_1)\,\Psi^3\Psi^1(y_2)\,\Psi^1\Psi^2(y_3)\,
S^{---}(z) \,S^{---}(\overline z)\,S^{---}(w)\, S^{---}(\overline w)
\rangle
\\
&~~
\langle
\ee^{\ii p_1\cdot X(y_1)}\,\ee^{\ii p_2\cdot X(y_2)}\,\ee^{\ii p_3\cdot X(y_3)}
\rangle
\Bigr\}
\\
~~=~&
\frac{g}{4\pi^4}\,{\rm Tr}\Big(\det C\,
F^{01}(p_1) F^{13}(p_2)
F^{30}(p_3))~\times
\\
&\int\frac{\prod_i dy_i\, dz d\overline z
\,dw d\overline w}{dV_{\rm
CKG}}~
\frac{(y_1-y_2)(y_1-y_3)(y_2-y_3)(z-\overline z)(w-\overline w)}{\prod_i
(y_i - z)(y_i - \overline z)(y_i - w)(y_i - \overline w)}
\end{aligned}
\end{equation}
where in the last step we have understood, as usual, the
$\delta$-function of momentum conservation. We now exploit the
$\mathrm{Sl}(2,\mathbb{R})$ invariance to fix
$y_1\to\infty$, $z\to \ii$ and $\overline z \to -\ii$, so that the
integrals in (\ref{FFFRRRR1}) become
\begin{equation}
\label{integral1}
4\,\int_{-\infty}^{+\infty}\!\!\!dy_2
\int_{-\infty}^{y_2}\!\!\!dy_3\,\,\frac{(y_2-y_3)}
{(y_2^2+1)(y_3^2+1)}\,
\int_{{\rm Im} w\geq 0}\!\!\! dw d\overline w\,
\frac{w- \overline w}{(y_2-w)(y_2-\overline
w)(y_3-w)(y_3-\overline w)}~~.
\end{equation}
Using the result (\ref{integralfin}) of appendix \ref{subsec:useful_integral},
Eq. (\ref{integral1}) reduces to the integrals (\ref{integral}) and yields, as final result, just a factor
of $4\pi^4$. Thus, the amplitude (\ref{FFFRRRR1}) gives rise to
the term $ -g\,{\rm Tr}\big(\det C\,
F^{01} F^{13} F^{30}\big)$ in the effective Lagrangian,
which easily generalizes to
\begin{equation}
-\frac{g}{3}\sum_{I\not = J \not = K}\!{\rm
Tr}\Big(
\det C\,F^{IJ}F^{JK}F^{KI}
\Big)
\label{holosupot2}
\end{equation}
{\it i.e.} the last term of (\ref{holsuppot}).

We conclude our analysis with a few general comments. All amplitudes we
have computed in the presence of the R-R background involve the
evaluation of some integrals over the world-sheet variables even after fixing the
$\mathrm{SL}(2,\mathbb{R})$ invariance, since they contain correlation
functions among more than three vertex operators.
Therefore, before ascribing the final result of these amplitudes to the effective
field theory action, one should study their factorization properties
in order to distinguish among  possible exchange contributions
and select the irreducible ones. In the case of amplitudes with
just a single insertion of the R-R graviphoton vertex
(\ref{RRvop}), it is quite easy to realize that these amplitudes could
be factorized only in an open string channel. However,
the intermediate states which would be exchanged in such a channel
are massive, since no coupling among massless states could give
rise to these exchange diagrams. Amplitudes which can be factorized only on
massive modes do not correspond to exchange diagrams in the effective field
theory, but rather to contact interactions. This is precisely
the case of the various amplitudes with a single closed string insertion
that we discussed in sections \ref{subsec:gauge_sector} and
\ref{subsec:superpotential_sector}. Things could be different, however,
for the last amplitude (\ref{FFFRRRR0}) which has two R-R
insertions, and hence can be factorized also in a closed string
channel as indicated in Fig. \ref{fig:factorization}.
\FIGURE{
\psfrag{F10}{$F^{10}$}
\psfrag{F03}{$F^{03}$}
\psfrag{F31}{$F^{31}$}
\psfrag{Cab}{\small $C_{\alpha\beta}$}
\psfrag{Cgd}{\small $C_{\gamma\delta}$}
\psfrag{to}{$\longrightarrow$}
\centerline{
\includegraphics[width=0.7\textwidth]{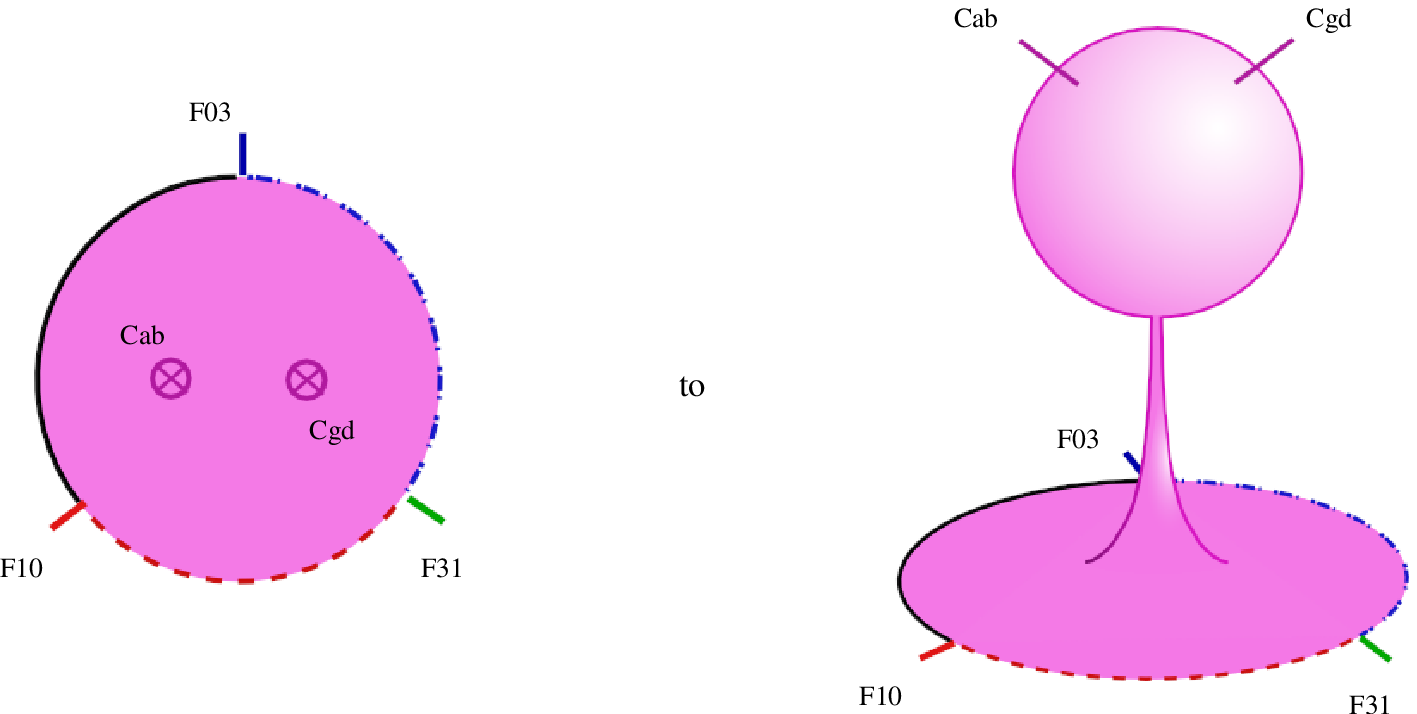}%
\caption{Factorization of the amplitude $\lvev V_{F^{01}} V_{F^{13}} V_{F^{30}}
V_C\,V_C\rvev$ in the closed channel.}
\label{fig:factorization}
}}

However, taking into account the explicit form of the R-R vertices (\ref{RRvop}),
it is quite easy to realize that also in this case the
exchanged closed string state is massive. In fact it corresponds to the
following NS-NS vertex operator
\begin{equation}
\label{Pvertex}
V(z,\overline z) \,\sim\,
\ee^{\ii(\varphi_1 + \varphi_2 + \varphi_3)(z)}\,
\ee^{-\phi(z)}\,\,
\ee^{\ii(\tilde\varphi_1 + \tilde\varphi_2 + \tilde\varphi_3)(\overline z)}
\,\ee^{-\tilde\phi(\overline z)}\,\,
\ee^{\ii p\cdot X(z,\overline z)}~~,
\end{equation}
which is physical when $p^2=-8$. Notice that in our diagram, the
momentum flowing in the intermediate channel is zero, since the
two external R-R vertices have both $p=0$, and thus in the
propagator of the virtual intermediate state only the mass term contributes.
Again, this is not an exchange diagram of the effective theory,
but rather a contact interaction, and thus the complete string
amplitude (\ref{FFFRRRR0}) in the limit $\alpha' \to 0$
must be assigned to the effective Lagrangian,
as we did. Notice that this is also consistent with the fact that in our
calculation we did not encounter any divergence, which, in presence of external
states at zero momentum, would be typically associated to the
exchange of some virtual massless particles.

\vskip 0.5cm
In conclusion we may say that our results prove
that the NAC deformation of gauge theories is
completely explained by the presence of a R-R graviphoton
background with constant (self-dual) field strength; this closed string
background modifies the open string dynamics
by introducing new types of
interactions that can be easily obtained by computing mixed
open/closed string amplitudes on disks with mixed boundary
conditions. In this paper we have explicitly considered the specific example of
the quiver theory corresponding to the orbifold
$\mathbb{C}^3/(\mathbb{Z}_2\times\mathbb{Z}_2)$, but since our
method is completely general, it could be applied to other orbifolds as
well, or to other configurations of D-branes, like for example D-branes
at angles. Furthermore, this approach can be used to analyze the effects
produced by other types of closed string fluxes on the effective
dynamics of open strings.

\vskip 1cm
\noindent {\large {\bf Acknowledgments}}
\vskip 0.2cm
\noindent We thank Silvia Penati for several fruitful discussions.

\vspace{0.5cm}
\appendix
\section{Appendix}
\label{appendix}
In this appendix we collect our conventions and several technical
details that are useful for the calculations reported in the main
text.

\vskip 0.5cm
\subsection{Conventions}
\label{subsec:conventions}
The matrices $(\sigma^\mu)_{\alpha\dot\beta}$ and
$(\bar\sigma^{\mu})^{\dot\alpha\beta}$ are defined by
\begin{equation}
\label{sigmas}
\sigma^\mu =
(\ii\vec\tau,\mathbf{1})~~~,~~~
\bar\sigma^\mu =
\sigma_\mu^\dagger = (-\ii\vec\tau,\mathbf{1})~~,
\end{equation}
where $\vec\tau$ are the ordinary Pauli matrices. They satisfy
the Clifford algebra
\begin{equation}
\label{cliff4}
\sigma_\mu\bar\sigma_\nu + \sigma_\nu\bar\sigma_\mu =
2\delta_{\mu\nu}\,\mathbf{1}~~,
\end{equation}
and correspond to a Weyl representation of the $\gamma$-matrices
acting on chiral or anti-chiral spinors $\psi_\alpha$ or $\psi^{\dot\alpha}$.
Out of these matrices, the $\mathrm{SO}(4)$ generators are defined by
\begin{equation}
\label{sigmamunu}
\sigma_{\mu\nu}
=\frac{1}{2}(\sigma_\mu\bar\sigma_\nu -
\sigma_\nu\bar\sigma_\mu)~~~,~~~
\bar\sigma_{\mu\nu}
=\frac{1}{2}(\bar\sigma_\mu\sigma_\nu -
\bar\sigma_\nu\sigma_\mu)~~.
\end{equation}
The matrices $\sigma_{\mu\nu}$ are self-dual and thus generate the
$\mathrm{SU}(2)_\mathrm{L}$ factor of $\mathrm{SO}(4)$;
the anti self-dual matrices $\bar\sigma_{\mu\nu}$ generate
instead the $\mathrm{SU}(2)_\mathrm{R}$ factor.
We raise and lower spinor indices as follows
\begin{equation}
\psi^\alpha=\varepsilon^{\alpha\beta}\,\psi_\beta
~~~,~~~
\psi_{\dot\alpha}= \varepsilon_{\dot\alpha\dot\beta}\,\psi^{\dot\beta}
\end{equation}
where $\varepsilon^{12}=\varepsilon_{12}
=-\varepsilon^{\dot 1\dot 2}=-\varepsilon_{\dot 1\dot 2}=+1$.
{F}rom these rules it follows that
\begin{equation}
\psi^\alpha\,\psi^\beta\,=\,-\frac{1}{2}\,\varepsilon^{\alpha\beta}\,
\psi\,\psi~~~,~~~
\bar\psi^{\dot\alpha}\,\bar\psi^{\dot\beta}\,
=\,-\frac{1}{2}\,\varepsilon^{\dot\alpha\dot\beta}\,
\bar\psi\,\bar\psi
\label{psipsi}
\end{equation}
and
\begin{equation}
\psi\,\sigma^\mu\bar\psi\,\,\psi\,\sigma^\nu\bar\psi
\,=\,\frac{1}{2}\,\psi\psi\,\bar\psi\bar\psi\,\delta^{\mu\nu}
\end{equation}

\vskip 0.5cm
\subsection{World-sheet correlation functions}
\label{subsec:correlators}
We report here some correlation functions among conformal fields
that are needed for the calculation of the string amplitudes of sections
\ref{sec:frac} and \ref{sec:def}.

\paragraph{Space-time correlators}
\begin{equation}
\label{curr2spinso4}
\langle
:\!\psi^\mu\psi^\nu\!:\!(y) \,S_{\alpha}(z)\, S_{\beta}(\overline z)
\rangle =
\frac{1}{2} \,
(\sigma^{\mu\nu})_{\alpha\beta}
(z-\overline z)^{\frac{1}{2}}\,(y-z)^{-1}\,(y-\overline z)^{-1}~~.
\end{equation}
\begin{equation}
\label{correlatorRM}
\langle \psi^{\mu}(y_1) \, \psi^{\nu}(y_2)\, S_{\alpha}(z) \,
S_{\beta}(\overline z) \rangle =
A(y_1, y_2, z, \overline z)\,\delta^{\mu\nu}\,\varepsilon_{\alpha\beta} +
B(y_1, y_2, z, \overline z) (\sigma^{\mu\nu})_{\alpha\beta}
\end{equation}
where
\begin{equation}
\label{A}
A(y_1, y_2, z, \overline z) = \frac{1}{2}\,
\frac{(y_1-z)(y_2-\overline z)+(y_2-z)(y_1-\overline z)}{(y_1-y_2)
\big[(y_1-z)(y_1-\overline z)(y_2-z)(y_2-\overline z)(z-\overline z)\big]^{\frac{1}{2}}}
\end{equation}
and
\begin{equation}
\label{B}
B(y_1, y_2, z, \overline z) = -\frac{1}{2}\,
\Bigg[\frac{(z-\overline z)}{(y_1-z)(y_1-\overline z)(y_2-z)(y_2-\overline z)}
\Bigg]^{\frac{1}{2}}~~.
\end{equation}
\begin{equation}
\label{4S_correlator}
\begin{aligned}
\langle S_{\gamma}(y_2)\, S_{\delta}(y_3)& \,
S_{\alpha}(z)\, S_{\beta}(\overline z) \rangle =
\big[{\varepsilon_{\gamma\delta}\,\varepsilon_{\alpha\beta}\,(y_2-\overline z)
(y_3-z)-\varepsilon_{\gamma\beta}\,
\varepsilon_{\delta\alpha}\,(y_2-y_3)(z-\overline z)}\big]
\\
&\,\times\,{\big[(y_2-y_3)(y_2-z)(y_2-\overline z)(y_3-z)
(y_3-\overline z)(z-\overline z)\big]^{-\frac{1}{2}}}~~.
\end{aligned}
\end{equation}

\paragraph{Internal space correlators}
\begin{equation}
\langle\Psi^1(y_1)\,\Psi^2\Psi^3(y_2)
\,S^{---}(z) \,S^{---}(\overline z)\rangle
= |y_1-z|^{-1}\,|y_2-z|^{-2}\,(z-\overline z)^{\frac{3}{4}}~~.
\end{equation}
\begin{equation}
\begin{aligned}
\langle
\Psi^2\Psi^3(y_1)\,\Psi^3\Psi^1(y_2)\,\Psi^1\Psi^2(y_3)\,
S^{---}(z_1) \,S^{---}(\overline z_1)\,S^{---}(z_2)\, S^{---}(\overline z_2)
\rangle \\
= \prod_{i=1}^3\prod_{a=1}^2 |y_i-z_a|^{-2}\,
\,\prod_{i<j}(y_i-y_j) \,\,\prod_{a<b}|z_a-z_b|^{\frac{3}{2}}
\,\,\prod_{a,b}|z_a-\overline z_b|^{\frac{3}{2}}~~.
\end{aligned}
\end{equation}

\paragraph{Superghost correlators}
\begin{equation}
\label{3sghostcorr}
\langle\ee^{-\phi(y_1)}\,\ee^{-\frac{1}{2}\phi(y_2)}\,
\ee^{-\frac{1}{2}\phi(y_3)}\rangle
= (y_1-y_2)^{-\frac{1}{2}}\,(y_1-y_3)^{-\frac{1}{2}}\,
(y_2-y_3)^{-\frac{1}{4}}~~,
\end{equation}
\begin{equation}
\label{4sghostcorr}
\begin{aligned}
{}&\!\!\!\!
\langle\ee^{-\frac{1}{2}\phi(y_1)}\,\ee^{-\frac{1}{2}\phi(y_2)}\,
\ee^{-\frac{1}{2}\phi(y_3)}\,
\ee^{-\frac{1}{2}\phi(y_4)}\rangle
\\
{}&~~= \Big[(y_1-y_2)\,(y_1-y_3)\,(y_1-y_4)\,(y_2 - y_3)\,
(y_2 - y_4)\,(y_3 - y_4)\Big]^{-\frac{1}{4}}~~.
\end{aligned}
\end{equation}

\vskip 0.5cm
\subsection{A useful integral}
\label{subsec:useful_integral}
In the calculation of the string amplitude $\lvev V_{F^{01}} V_{F^{13}} V_{F^{30}}
V_C\,V_C\rvev$, one of the
ingredients is the following integral in the upper half complex plane $\mathcal{H}_+$
\begin{equation}
\label{intz1}
I(a,b) = \int_{\mathcal{H}_+} \!\!\!dw \,d\overline w\,\,
\frac{(w - \overline w)}{(w - a)(\overline w -a)(w-b)
(\overline w -b)}
\end{equation}
with $(a,b\in \mathbb{R}~~,~~a>b)$ (see Eq. (\ref{integral1})).
As it stands, the integral $I(a,b)$ is formally divergent. We may regularize
it by excluding the real axis from the integration region, {\it i.e.} we take
${\rm Im}\,w\geq \epsilon$, where $\epsilon$ will be sent to zero at the end of the
calculation. Notice that this regularization prescription is
precisely what is required in mixed open/closed string amplitudes.
With this in mind, after applying Stoke's theorem, we have
\begin{equation}
\label{intz2}
I(a,b;\epsilon) =  \oint_{\partial\mathcal{H}_{+\epsilon}}
\!\!\!d\overline w\,\,\frac{\ln \left[(w - a)(w-b)\right]}{2(\overline w -
a)(\overline w - b)}
+\oint_{\partial\mathcal{H}_{+\epsilon}}
\!\!\!dw\,\,\frac{\ln \left[(\overline w - a)(\overline w-b)\right]}{2(w -
a)(w - b)}~~.
\end{equation}
On the integration path, we have
\begin{equation}
w=x+\ii \epsilon~~~~,~~~~\overline w = x - \ii
\epsilon~~~~,~~~~dw=d\overline w=dx
\end{equation}
with $-\infty < x < +\infty$. Thus, the first integral in
(\ref{intz2}) becomes
\begin{equation}
I_1(a,b;\epsilon) = \int_{-\infty}^{+\infty}\!\!\!
dx\,\,\frac{\ln \left[(x - a+\ii \epsilon)(x-b+\ii \epsilon)\right]}{2(x -
a-\ii \epsilon)(x - b-\ii \epsilon)}
\end{equation}
which can be easily evaluated using Jordan's lemma and residues
theorem. In fact, we get
\begin{equation}
\label{intz3}
\begin{aligned}
I_1(a,b;\epsilon)
&=2\pi\ii\Bigg\{\frac{\ln\big[(2\ii\epsilon)(a-b+2\ii\epsilon)\big]}{2(a-b)}
+\frac{\ln\big[(b-a+2\ii\epsilon)(2\ii\epsilon)\big]}{2(b-a)}\Bigg\}
\\
&=\frac{\pi\ii}{a-b}\Big[\ln(a-b+2\ii\epsilon)-\ln(b-a+2\ii\epsilon)\Big]~~.
\end{aligned}
\end{equation}
Now we can safely take the limit $\epsilon\to 0^+$, and using the
fact that
\begin{equation}
\lim_{\epsilon\to
0^+}\Big[\ln(a-b+2\ii\epsilon)-\ln(b-a+2\ii\epsilon)\Big]=
\pi\ii~~,
\end{equation}
we finally get
\begin{equation}
I_1(a,b) = \frac{\pi^2}{a-b}~~.
\label{integralfin0}
\end{equation}
The calculation of the second integral in (\ref{intz2}) proceeds
along the same lines and yields the same result,
\begin{equation}
I_2(a,b) = \frac{\pi^2}{a-b}~~;
\label{integralfin00}
\end{equation}
thus in the end we have
\begin{equation}
I(a,b) = I_1(a,b)+I_2(a,b)=\frac{2\pi^2}{a-b}~~.
\label{integralfin}
\end{equation}


\vspace{0.5cm}

\begin{thebibliography}{99}

\bibitem{Seiberg:1999vs}
V.~Schomerus,
%``D-branes and deformation quantization,''
JHEP {\bf 9906} (1999) 030
[arXiv:hep-th/9903205].
%%CITATION = HEP-TH 9903205;%%
N.~Seiberg and E.~Witten,
%``String theory and noncommutative geometry,''
JHEP {\bf 9909} (1999) 032
[arXiv:hep-th/9908142].
%%CITATION = HEP-TH 9908142;%%

\bibitem{deBoer:2003dn}
J.~de Boer, P.~A.~Grassi and P.~van Nieuwenhuizen,
%``Non-commutative superspace from string theory,''
Phys.\ Lett.\ B {\bf 574} (2003) 98
[arXiv:hep-th/0302078].
%%CITATION = HEP-TH 0302078;%%

\bibitem{Ooguri:2003qp}
H.~Ooguri and C.~Vafa,
%``The C-deformation of gluino and non-planar diagrams,''
Adv.\ Theor.\ Math.\ Phys.\  {\bf 7} (2003) 53
[arXiv:hep-th/0302109].
%%CITATION = HEP-TH 0302109;%%

\bibitem{Ooguri:2003tt}
H.~Ooguri and C.~Vafa,
``Gravity induced C-deformation,''
[arXiv:hep-th/0303063].
%%CITATION = HEP-TH 0303063;%%

\bibitem{Seiberg:2003yz}
N.~Seiberg,
%``Noncommutative superspace, N = 1/2 supersymmetry, field theory and  string theory,''
JHEP {\bf 0306} (2003) 010
[arXiv:hep-th/0305248].
%%CITATION = HEP-TH 0305248;%%

\bibitem{Berkovits:2003kj}
N.~Berkovits and N.~Seiberg,
%``Superstrings in graviphoton background and N = 1/2 + 3/2 supersymmetry,''
JHEP {\bf 0307} (2003) 010
[arXiv:hep-th/0306226].
%%CITATION = HEP-TH 0306226;%%

\bibitem{Ferrara:2000mm}
S.~Ferrara and M.~A.~Lledo,
%``Some aspects of deformations of supersymmetric field theories,''
JHEP {\bf 0005} (2000) 008
[arXiv:hep-th/0002084].
%%CITATION = HEP-TH 0002084;%%

\bibitem{Klemm:2001yu}
D.~Klemm, S.~Penati and L.~Tamassia,
%``Non(anti)commutative superspace,''
Class.\ Quant.\ Grav.\  {\bf 20} (2003) 2905
[arXiv:hep-th/0104190].
%%CITATION = HEP-TH 0104190;%%

\bibitem{Britto:2003aj}
R.~Britto, B.~Feng and S.~J.~Rey,
 %``Deformed superspace, N = 1/2 supersymmetry and (non)renormalization
%theorems,''
JHEP {\bf 0307} (2003) 067
[arXiv:hep-th/0306215];
%%CITATION = HEP-TH 0306215;%%
R.~Britto, B.~Feng and S.~J.~Rey,
%``Non(anti)commutative superspace, UV/IR mixing and open Wilson lines,''
JHEP {\bf 0308} (2003) 001
[arXiv:hep-th/0307091].
%%CITATION = HEP-TH 0307091;%%

\bibitem{Terashima:2003ri}
S.~Terashima and J.~T.~Yee,
%``Comments on noncommutative superspace,''
JHEP {\bf 0312} (2003) 053
[arXiv:hep-th/0306237].
%%CITATION = HEP-TH 0306237;%%

\bibitem{Araki:2003se}
T.~Araki, K.~Ito and A.~Ohtsuka,
%``Supersymmetric gauge theories on noncommutative superspace,''
Phys.\ Lett.\ B {\bf 573} (2003) 209
[arXiv:hep-th/0307076].
%%CITATION = HEP-TH 0307076;%%

%\cite{Chandrasekhar:2003uq}
\bibitem{Chandrasekhar:2003uq}
  B.~Chandrasekhar and A.~Kumar,
  %``D = 2, N = 2, supersymmetric theories on non(anti)commutative
  %superspace,''
  JHEP {\bf 0403} (2004) 013
  [arXiv:hep-th/0310137].
  %%CITATION = HEP-TH 0310137;%%

\bibitem{Grisaru:2003fd}
M.~T.~Grisaru, S.~Penati and A.~Romagnoni,
 %``Two-loop renormalization for nonanticommutative N = 1/2 supersymmetric WZ
%model,''
JHEP {\bf 0308} (2003) 003
[arXiv:hep-th/0307099];
%%CITATION = HEP-TH 0307099;%%
A.~Romagnoni,
%``Renormalizability of N = 1/2 Wess-Zumino model in superspace,''
JHEP {\bf 0310} (2003) 016
[arXiv:hep-th/0307209].
%%CITATION = HEP-TH 0307209;%%

\bibitem{Britto:2003kg}
R.~Britto and B.~Feng,
%``N = 1/2 Wess-Zumino model is renormalizable,''
Phys.\ Rev.\ Lett.\  {\bf 91} (2003) 201601
[arXiv:hep-th/0307165].
%%CITATION = HEP-TH 0307165;%%

\bibitem{Lunin:2003bm}
O.~Lunin and S.~J.~Rey,
 %``Renormalizability of non(anti)commutative gauge theories with N = 1/2
%supersymmetry,''
JHEP {\bf 0309} (2003) 045
[arXiv:hep-th/0307275].
%%CITATION = HEP-TH 0307275;%%

\bibitem{Berenstein:2003sr}
D.~Berenstein and S.~J.~Rey,
 %``Wilsonian proof for renormalizability of N = 1/2 supersymmetric field
%theories,''
Phys.\ Rev.\ D {\bf 68} (2003) 121701
[arXiv:hep-th/0308049].
%%CITATION = HEP-TH 0308049;%%

\bibitem{Alishahiha:2003kg}
M.~Alishahiha, A.~Ghodsi and N.~Sadooghi,
``One-loop perturbative corrections to non(anti)commutativity parameter of N =
1/2 supersymmetric U(N) gauge theory,''
[arXiv:hep-th/0309037].
%%CITATION = HEP-TH 0309037;%%

\bibitem{Imaanpur:2003jj}
A.~Imaanpur,
%``On instantons and zero modes of N = 1/2 SYM theory,''
JHEP {\bf 0309} (2003) 077
[arXiv:hep-th/0308171];
%%CITATION = HEP-TH 0308171;%%
A.~Imaanpur,
%``Comments on gluino condensates in N = 1/2 SYM theory,''
JHEP {\bf 0312} (2003) 009
[arXiv:hep-th/0311137].
%%CITATION = HEP-TH 0311137;%%

\bibitem{Grassi:2003qk}
P.~A.~Grassi, R.~Ricci and D.~Robles-Llana,
``Instanton calculations for N = 1/2 super Yang-Mills theory,''
[arXiv:hep-th/0311155].
%%CITATION = HEP-TH 0311155;%%

\bibitem{Britto:2003uv}
R.~Britto, B.~Feng, O.~Lunin and S.~J.~Rey,
``U(N) instantons on N = 1/2 superspace: Exact solution and geometry of moduli
space,''
[arXiv:hep-th/0311275].
%%CITATION = HEP-TH 0311275;%%

%\cite{Banin:2004hy}
\bibitem{Banin:2004hy}
A.~T.~Banin, I.~L.~Buchbinder and N.~G.~Pletnev,
%``Chiral effective potential in N = 1/2 non-commutative Wess-Zumino model,''
JHEP {\bf 0407}, 011 (2004)
[arXiv:hep-th/0405063].
%%CITATION = HEP-TH 0405063;%%

%\cite{Jack:2004pq}
\bibitem{Jack:2004pq}
I.~Jack, D.~R.~T.~Jones and L.~A.~Worthy,
%``One-loop renormalisation of N = 1/2 supersymmetric gauge theory,''
arXiv:hep-th/0412009.
%%CITATION = HEP-TH 0412009;%%

%\cite{Penati:2004gh}
\bibitem{Penati:2004gh}
S.~Penati and A.~Romagnoni,
%``Covariant quantization of N = 1/2 SYM theories and supergauge invariance,''
arXiv:hep-th/0412041.
%%CITATION = HEP-TH 0412041;%%

%\cite{Azorkina:2005mx}
\bibitem{Azorkina:2005mx}
O.~D.~Azorkina, A.~T.~Banin, I.~L.~Buchbinder and N.~G.~Pletnev,
%``Generic chiral superfield model on nonanticommutative N = 1/2 superspace,''
arXiv:hep-th/0502008.
%%CITATION = HEP-TH 0502008;%%

%\cite{Hatanaka:2005rg}
\bibitem{Hatanaka:2005rg}
T.~Hatanaka, S.~V.~Ketov, Y.~Kobayashi and S.~Sasaki,
%``Non-anti-commutative deformation of effective potentials in supersymmetric
%gauge theories,''
arXiv:hep-th/0502026.

\bibitem{Billo:2004zq}
M.~Billo, M.~Frau, I.~Pesando and A.~Lerda,
%``N = 1/2 gauge theory and its instanton moduli space from open strings in R-R
%background,''
JHEP {\bf 0405} (2004) 023
[arXiv:hep-th/0402160].
%%CITATION = HEP-TH 0402160;%%

\bibitem{quivers}
M.~R.~Douglas and G.~W.~Moore,
%``D-branes, Quivers, and ALE Instantons,''
arXiv:hep-th/9603167;
%%CITATION = HEP-TH 9603167;%%

\bibitem{fractional}
M.~R.~Douglas,
%``Enhanced gauge symmetry in M(atrix) theory,''
JHEP {\bf 9707} (1997) 004
[arXiv:hep-th/9612126];
%%CITATION = HEP-TH 9612126;%%
D.~E.~Diaconescu, M.~R.~Douglas and J.~Gomis,
%``Fractional branes and wrapped branes,''
JHEP {\bf 9802} (1998) 013
[arXiv:hep-th/9712230];
%%CITATION = HEP-TH 9712230;%%
D.~E.~Diaconescu and J.~Gomis,
%``Fractional branes and boundary states in orbifold theories,''
JHEP {\bf 0010} (2000) 001
[arXiv:hep-th/9906242];
%%CITATION = HEP-TH 9906242;%%
M.~R.~Gaberdiel and B.~J.~Stefanski,
%``Dirichlet branes on orbifolds,''
Nucl.\ Phys.\ B {\bf 578} (2000) 58
[arXiv:hep-th/9910109];
%%CITATION = HEP-TH 9910109;%%
T.~Takayanagi,
%``String creation and monodromy from fractional D-branes on ALE spaces,''
JHEP {\bf 0002} (2000) 040
[arXiv:hep-th/9912157];
%%CITATION = HEP-TH 9912157;%%
M.~Billo, B.~Craps and F.~Roose,
%``Orbifold boundary states from Cardy's condition,''
JHEP {\bf 0101} (2001) 038
[arXiv:hep-th/0011060].
%%CITATION = HEP-TH 0011060;%%

\bibitem{FID_string}
M.~Dine, N.~Seiberg and E.~Witten,
%``Fayet-Iliopoulos Terms In String Theory,''
Nucl.\ Phys.\ B {\bf 289}, 589 (1987);
%%CITATION = NUPHA,B289,589;%%
J.~J.~Atick, L.~J.~Dixon and A.~Sen,
%``String Calculation Of Fayet-Iliopoulos D Terms In Arbitrary Supersymmetric
%Compactifications,''
Nucl.\ Phys.\ B {\bf 292}, 109 (1987);
%%CITATION = NUPHA,B292,109;%%
M.~Dine, I.~Ichinose and N.~Seiberg,
%``F Terms And D Terms In String Theory,''
Nucl.\ Phys.\ B {\bf 293}, 253 (1987).
%%CITATION = NUPHA,B293,253;%%

%\cite{Imaanpur:2005pd}
\bibitem{Imaanpur:2005pd}
A.~Imaanpur,
%``Supersymmetric D3-branes in five-form flux,''
arXiv:hep-th/0501167.
%%CITATION = HEP-TH 0501167;%%

\bibitem{Billo:2002hm}
M.~Billo, M.~Frau, I.~Pesando, F.~Fucito, A.~Lerda and A.~Liccardo,
%``Classical gauge instantons from open strings,''
JHEP {\bf 0302} (2003) 045
[arXiv:hep-th/0211250].
%%CITATION = HEP-TH 0211250;%%


\end{thebibliography}
\end{document}